\documentclass[aps,prb,floatfix,twocolumn,notitlepage,superscriptaddress,10pt]{revtex4-2}
\usepackage{xcolor}
\usepackage{grffile}
\usepackage{amsmath, amsthm, amssymb, bbold}
\usepackage[normalem]{ulem}
\usepackage{cancel}
\usepackage{bm}
\usepackage{microtype}
\usepackage{hyperref}
\usepackage{setspace}
\usepackage{grffile}
\hypersetup{colorlinks,linkcolor=blue,urlcolor=blue,citecolor=blue}
\usepackage{amsmath}    % need for subequations
\usepackage{amssymb}
\usepackage{graphicx}   % need for figures
\usepackage{verbatim}   % useful for program listings
\usepackage{color}      % use if color is used in text
\usepackage{hyperref}   % use for hypertext links, including those to external documents and URLs
\usepackage[normalem]{ulem}
\usepackage{natbib}
\usepackage{enumitem} 
\usepackage{dsfont}

\newcommand{\gzb}[1]{  {\color{blue} #1} }
\newcommand{\beq}{\begin{equation}}
\newcommand{\eeq}{\end{equation}}
\newcommand{\bea}{\begin{eqnarray}}
\newcommand{\eea}{\end{eqnarray}}

\newcommand{\be}{\begin{equation}}
\newcommand{\ee}{\end{equation}}

\definecolor{darkgreen}{rgb}{0,0.5,0}
\definecolor{orange}{rgb}{1,0.5,0}
\definecolor{grey}{rgb}{.6,.6,.6}

\newcommand{\bra}[1]{\langle #1|}
\newcommand{\ket}[1]{|#1\rangle}
\newcommand{\average}[1]{\langle #1\rangle}

\newcommand{\bS}{\mathbf S}

\newcommand{\cL}{{\cal L }}

\newcommand{\cD}{{\cal D}}

\newcommand{\ketL}[1]{|#1 )}

\newcommand{\tp}[1]{{\color{darkgreen} #1}}

\begin{document}
\title{Kardar-Parisi-Zhang scaling in the  Hubbard model}
 \author{C\u at\u alin Pa\c scu Moca}
\affiliation{Department of Theoretical Physics, Institute of Physics, Budapest University of Technology and Economics, Budafoki \'ut 8., H-1111 Budapest, Hungary}
\affiliation{Department of Physics, University of Oradea,  410087, Oradea, Romania}
\author{Mikl\'os Antal Werner}
\affiliation{Department of Theoretical Physics, Institute of Physics, Budapest University of Technology and Economics, Budafoki \'ut 8., H-1111 Budapest, Hungary}
\affiliation{MTA-BME Quantum Dynamics and Correlations Research Group, 
Institute of Physics, Budapest University of Technology and Economics,  Budafoki \'ut 8., H-1111 Budapest, Hungary}
\affiliation{Strongly Correlated Systems Lend\"ulet Research Group,
Wigner Research Centre for Physics, H-1525, Budapest, Hungary}
\author{Angelo Valli}
\affiliation{Department of Theoretical Physics, Institute of Physics, Budapest University of Technology and Economics, Budafoki \'ut 8., H-1111 Budapest, Hungary}
% \author{\" Ors Legeza}
% \affiliation{Strongly Correlated Systems Lend\"ulet Research Group,
% Wigner Research Centre for Physics, H-1525, Budapest, Hungary}
% %\affiliation{ Fachbereich Physik, Philipps-Universit\"at Marburg, 35032 Marburg, Germany}	
% \affiliation{Institute for Advanced Study,Technical University of Munich, Lichtenbergstrasse 2a, 85748 Garching, Germany}
%\affiliation{MTA-BME Quantum-Dynamics and Correlations Research Group, E\"otvo\"s Lor\'and Research Network (ELKH), Budapest University of Technology and Economics, 1111 Budapest, Budafoki \'ut 8, Hungary}
\author{Gergely Zar\'and}
\affiliation{Department of Theoretical Physics,  Institute of Physics, Budapest University of Technology and Economics,  Budafoki \'ut 8., H-1111 Budapest, Hungary}
\affiliation{MTA-BME Quantum Dynamics and Correlations Research Group, 
Institute of Physics, Budapest University of Technology and Economics,  Budafoki \'ut 8., H-1111 Budapest, Hungary}
%\affiliation{ BME-MTA Exotic Quantum Phases Group, Institute of Physics, 
%Budapest University of Technology and Economics, H-1111 Budapest, Hungary}
\author{Toma\v z Prosen}
\affiliation{Department of Physics, Faculty of Mathematics and Physics,
University of Ljubljana, Jadranska 19, SI-1000 Ljubljana, Slovenia}	 
\date{\today}

\begin{abstract}
 We explore the Kardar-Parisi-Zhang (KPZ) scaling in the one-dimensional Hubbard model, which exhibits global $SU_c(2)\otimes SU_s(2)$ symmetry at half-filling, for the pseudo-charge and the total spin.
 We analyze dynamical scaling properties of  high temperature charge and spin correlations and transport.
% \gz{of the high temperature structure factor \bf I am not sure what we mean by this} for the charge and magnetization,
% as well as the associated charge and spin currents. 
At half-filling, we observe a clear KPZ scaling in both charge and spin sectors.
 Away from half-filling, the $SU_c(2)$ charge symmetry is reduced to $U_c(1)$, while the $SU_s(2)$ symmetry for the total spin is retained. Consequently,  transport in the charge sector becomes ballistic, while  KPZ scaling is preserved in the spin sector. These findings confirm the link between non-abelian symmetries and KPZ scaling in the presence of integrability. We study two settings of the model: one involving a quench from a bi-partitioned state asymptotically close to the $T=\infty$ equilibrium state of the system, and another where the system is coupled to two markovian reservoirs at the two edges of the chain.
\end{abstract}
\maketitle

%\gz{\bf If Pascu used indeed non-abelian symmetries, then we should offer co-authorship to Ors Legeza. }

\section{Introduction}

Universality, a crucial concept in statistical physics, refers to the similar behavior exhibited by diverse physical systems, despite having different microscopic details. This allows us to predict the behavior of complex physical systems based on a few universal characteristics instead of relying on the specifics of each system~\cite{Hinrichsen.2000,Geza.2004}.
In this context, originally, the KPZ (Kardar-Parisi-Zhang) universality class refers to a broad range of classical stochastic growth models that exhibit similar scaling behavior to the original KPZ equation~\cite{KPZ.86,Krug.1997,Kriecherbauer.2010,Corwin.2012}.
These models describe the evolution of interfaces between two media, such as the growth of a crystal or the motion of a fluid~\cite{Villain.1991,Timothy.1995}.
The key insight of the KPZ scaling is that the fluctuations at the interface exhibit a self-similar behavior, such that their statistical properties  are invariant under rescaling of time and space.

The KPZ scaling has also been shown to describe high-temperature dynamics of certain many-body systems near equilibrium~\cite{Marko.2019}, building on previous observations~\cite{Znidaric.2011,Zunkovic2013,Ljubotina.2017}.
This opened up new avenues for understanding the behavior of quantum many-body systems~\cite{Bertini.2021}, allowed to identify commonalities between classical and quantum systems~\cite{Zunkovic2013,Dhar2019,KP2020,roy2022robustness}, 
and provided a new framework for describing the complex dynamics of quantum quenches~\cite{Marko.2019}.
Through extensive investigations of near equilibrium dynamics of magnetization and spin currents in XXZ spin chains, it has been found that, exactly at the $SU(2)$ symmetric fixed point
(i.e., associated with no anisotropy of the spin interaction) the dynamical spin structure factor can be exactly described by the KPZ correlation function, and thus exhibits superdiffusive behavior with dynamical scaling exponent $z=3/2$. However, away from this point, the system exhibits either ballistic or diffusive behavior~\cite{Ljubotina_2017}.  Extensions to higher spin integrable models has 
been discussed as well~\cite{Krajnik2020,Ye.2022}.
It has thus been conjectured~\cite{Krajnik2020} that high-temperature dynamical two-point correlation functions of Noether charges of all integrable systems with non-abelian symmetries are described by the Pr\" ahofer-Spohn~\cite{Prahofer2004} scaling function of KPZ universality class. The numerical \cite{Bulchandani.2019,Dupont.2020,Schmitteckert.2020, Diessel.2022,Oliveira.2023,nandy2022spin} and experimental~\cite{Scheie.2021, Fontaine.2022, Wei.2022} evidence for the above conjecture has been mounting, whereas the proof or precise mechanism for its validity are still lacking.
There has been, however, a clear self-consistent explanation of anomalous scaling exponent $z=3/2$ within the framework of generalized hydrodynamics~\cite{Vasseur2019,Ilievski2021} (see also review~\cite{Bulchandani_2021}). Nevertheless, there remains a crucial distinction between KPZ scaling in non-abelian integrable spin chains, and classical KPZ universality e.g. in surface growth phenomena.
While the latter is clearly far from equilibrium and violating detailed balance, the former belongs to the domain of equilibrium physics. 
As such, distribution of fluctuations (i.e. full counting statistics) has to 
be symmetric and hence cannot be described by KPZ universality~\cite{KIP2022,DeNardis2022,KIP2023}. Therefore, 
more work is needed to understand even phenomenologically to what extent KPZ universality applies to non-abelian integrable spin chains.

So far, most studies have focused on spin models, and relatively less attention has been paid to non-abelian integrable fermionic models, such as the Hubbard model~\cite{Prosen.2012,Fava.2020}.
The Hubbard model is a widely studied model in condensed matter physics and describes interacting fermions on a
lattice with a local (on-site) repulsion~\cite{Imada.1998}. At general fillings, the Hubbard model has a global $U_c(1)\otimes SU_s(2)$ symmetry associated with the total charge $(c)$ and spin $(s)$ conservation. As we discuss it in detail in Sec.~\ref{sec:model}, the global symmetry is raised to $SU_c(2)\otimes SU_s(2)$ at half filling, which  makes the Hubbard model a particularly interesting candidate for investigating the connection between symmetries and KPZ scaling in fermionic systems. Additionally, the Hubbard model is integrable~\cite{Shastry.86}, which further adds to its appeal for such studies. The existence of two non-abelian symmetries in the model enables us to explore the conjecture in both the spin and charge sectors, thus providing us with a greater flexibility in examining the role of non-abelian symmetries.

We tackle the problem from two distinct setups. First, we study a quench protocol within a closed Hubbard chain by time evolving the density matrix $\rho(t)$. We consider a mixed initial state, asymptotically close to the infinite temperature ($T=\infty$) state, with a weak imbalance either in the occupation (charge density) or the magnetization across an interface in the middle of the chain. Within this setup, we determine the universal functions for the charge and magnetization gradients, as well as the charge and spin currents. Second, we examine an open system setup, where the Hubbard chain is locally coupled to external {markovian} reservoirs at both ends, and investigate how the charge current scales with system size in the non-equilibrium steady state (NESS).
Our results clearly corroborate the KPZ scaling conjecture. With respect to the first setup, our results are consistent with those of Ref.~\cite{Fava.2020}, while they represent an improvement with respect to convergence as well as extension to the regimes of partially broken non-abelian symmetries, i.e. considering initial states with non-half filling or non-zero magnetization, as well as offer some
quantitative studies of crossover regimes. With respect
to the second setup, our results are again consistent with KPZ scaling~\cite{Znidaric.2011}, while they are not consistent with the previous boundary driven Lindblad study of the Hubbard chain~\cite{Prosen.2012}. Our detailed study of convergence to NESS reported below shows that previous results~\cite{Prosen.2012} were not yet fully converged and hence displayed an 
illusion of Ohm's law behavior.

The structure of our paper is as follows: In Sec.~\ref{sec:model}, we provide an introduction to the Hubbard model and discuss its non-abelian symmetries. In Sec.~\ref{sec:quench}, we describe the quench protocol used to investigate the KPZ scaling. In Sec.~\ref{sec:ballistic}, we present the results for the non-interacting limit where transport is ballistic.
Moving on to the finite $U$ case, Sec.\ref{sec:superdiffusive} presents scaling results for the average occupation, average magnetization, and associated currents in the context of KPZ scaling. However, as we move away from half-filling in Sec.~\ref{sec:away_half_filling}, we show that the KPZ scaling is lost in the charge sector, and the system displays ballistic transport.
In Sec.~\ref{sec:wl}, we study the scaling of the NESS current in an open setup with respect to the system size. We show that it displays a superdiffusive behavior, thus corroborating the results obtained in the quench setup.
Finally, in Sec.~\ref{sec:conclusions}, we summarize our findings and present our conclusions.

\section{Hubbard model, symmetries and integrability}\label{sec:model}

The Hubbard model is a simplified model of the behavior of interacting fermions on a lattice, 
and has proven to be a valuable tool in understanding the properties of strongly correlated electron systems.
The Hamiltonian of the one-dimensional (1D) Hubbard chain on \tp{$L$} sites reads
\begin{equation}
  H = - { J\over 2} \sum_{\sigma}\sum_{x  ={-L/2}}^{L/2-2}\big ( c^\dagger_{x\sigma}c^{\phantom{\dagger}}_{x+1\sigma} +h.c.\big)
 + \frac U 2 \sum_{x}  % n_{x\uparrow}n_{x\downarrow},
 (n_x-1)^2
 \label{eq:Hubbard}
\end{equation}
where $c_{x\sigma}^{(\dagger)}$ denote the annihilation (creation) operators of a fermion at site $x$ on the chain, with spin $\sigma$,  and 
$n_{x}= \sum_\sigma c_{x\sigma}^{\dagger}c_{x\sigma}$ denotes the  occupation number operator.
%\gz{\bf G: One needs the chemical potential term to have the charge SU(2) symmetry, discussed later.}
The first term in the Hamiltonian describes the kinetic energy of the fermions, as they can hop between neighboring sites with amplitude $J$. The second term represents the Coulomb repulsion between fermions on the same site, where $U$ is the strength of the interaction. In our calculations we consider the energy units of $J$ and measure time in units of $1/J$, i.e. $J\equiv 1$.
In the Hubbard model, the local Hilbert space that refers to the set of possible states at a single lattice site has dimension $d=4$, and is spanned by the empty state, the two possible spin states of a fermion of spin $S=1/2$, and the doubly-occupied state.
Thus, the basis states of the local Hilbert space are denoted as $\{\ket{0}, \ket{\downarrow},
\ket{\uparrow}, \ket{\uparrow\downarrow} \}$. 

%\MAW{Away from half filling the modal has a global $U_c(1)\otimes SU_s(2)$ symmetry, and the} set of 4 states 
%can be organized in 3 multiplets~\cite{Moca.2022}. 
%\textcolor{teal}{AV: [there is a slight logic gap, as so far we only referred to $SU_c(2) \otimes SU_s(2)$ symmetry]}

The Hubbard model has several symmetries that are important for understanding its behavior~\cite{Grosse.1989}.
It exhibits an $SU_s(2)$ spin symmetry
%\footnote{Here the superscript $s$ simply denotes the spin channel of the SU(2) symmetry},
which arises from the total spin conservation, i.e., each component of
%, the 
%operator valued vector
\begin{equation}
  \bS = {1\over 2} \sum_{x,\sigma\sigma'} c_{x\sigma}^\dagger \boldsymbol{\sigma}_{\sigma\sigma'} c_{x\sigma'} ,
  \label{eq:S_tot}
\end{equation}  
commutes with the Hamiltonian~\eqref{eq:Hubbard}. % irrespective of the filling $\average{n}$. 
%%G I removed this, I think it does not make too much 
%It leads to interesting physical properties, 
%such as the existence of a spin gap and spin-charge separation in one-dimensional case~\cite{Voit.1993,Voit_1995,Guan.2013}. 
This symmetry is broken in the presence of an external magnetic field, or in the presence of a spin imbalance within the chain
($\average{n_{\uparrow}}\neq \average{n_{\downarrow}}$).

The Hubbard model displays also a $U_c(1)$ charge symmetry, that arises from the %local 
conservation of the total number of electrons in the system,
\begin{equation}
  N=\sum_{x,\sigma} c^\dagger_{x\sigma} c_{x\sigma}.
  \label{eq:charge} 
\end{equation}
%which in particular, it leads to the formation of charge density waves in certain regions of the phase diagram.
At half-filling,  $\langle n \rangle =1$ (i.e., $N=L$), the $U_c(1)$ symmetry is raised to a non-abelian $SU_c(2)$ symmetry,
sometimes referred to as  %'$\eta$-pairing $SU_c(2)$'~\cite{Moudgalya.2020}. 
 '$\eta$-pairing'~\cite{Moudgalya.2020}. The operators
$\eta^\dagger$ and $\eta$, defined as
\begin{equation}
\eta^\dagger =\sum_x (-1)^x c^\dagger_{x\uparrow} c^\dagger_{x\downarrow},\phantom{aa}
\eta =\sum_x (-1)^x  c_{x\downarrow}c_{x\uparrow},
\end{equation}  
together with the appropriately shifted particle number operator, $\eta_z \equiv N-L$, satisfy the $SU(2)$ algebra, and commute with the Hamiltonian of the half filled Hubbard model~\eqref{eq:Hubbard}, proving its $SU_c(2)$ symmetry, 
\begin{eqnarray}
  &[H,\eta^\dagger]=0 & \phantom{aa} [H,\eta]=0 \phantom{aa} [H,\eta_z] = 0 \label{eq:commutators}\\
  &[\eta_z, \eta^\dagger] = \eta^\dagger &  \phantom{aa}  [\eta_z, \eta] = -\eta  \phantom{aa}  [\eta^\dagger, \eta] = 2 \eta_z.  \nonumber
\end{eqnarray}
The square of the total pseudo-charge operator is
\begin{equation}
  \boldsymbol{\eta}^2 = {1\over 2}(\eta^\dagger \eta + \eta\, \eta^\dagger)+\eta_z^2,\label{eq:eta}
\end{equation}
and it also commutes with the Hamiltonian~\eqref{eq:Hubbard}. 

Time evolution with $H$ thus respects $SU_c(2)\times SU_s(2)$ symmetry. While the infinite temperature state also obeys these symmetries, 
more complex initial states  can, however, break them. At
the $SU_c(2)$ symmetric point the 4-dimensional local Hilbert space can be organized into two-dimensional charge-spin multiplets, $\lbrace \ket{\uparrow}, \ket{\downarrow} \rbrace$ and $\lbrace \ket{\uparrow \downarrow}, \ket{0} \rbrace$. Away from half filling, the second multiplet breaks into two one-dimensional states that are distinguished by their $U_c(1)$ charge~\cite{Moca.2022}. Similarly, creating a spin imbalance in the initial state 
breaks the $SU_s(2)$ symmetry. In the limit of very small symmetry breaking, however, linear response is still governed 
by the unperturbed, $SU_s(2)\times SU_c(2)$ symmetrical  system.

In addition to continuous symmetries, the Hubbard model with nearest-neighbor hopping displays also  other, discrete
symmetries such as the  particle-hole (p-h)~\cite{Denteneer.2001} or the duality symmetry~\cite{Carlstrom_2017,JAKUBCZYK2020293}, but in our analysis, only the continuous
abelian and non-abelian symmetries are relevant,   discrete symmetries have just been mentioned for the sake of completeness.
%\gz{\bf I found several typos in the text, a final spell-check would be useful before submission.}

%The concept of integrability is fundamental in physics. Integrable models are of great interest because they allow for a 
%deeper understanding of the underlying physical phenomena and for exact predictions that can be compared with experimental results.
%Integrability is often associated with an existence of infinitely-dimensional symmetry, and hence an 
%infinite (extensive) number of local or quasi-local conserved quantities, and the corresponding conservation laws including currents of conserved quantities.
%In physical systems, these quantities are often related to the conservation of energy or momentum  and can be used to derive exact solutions.

In addition to the global symmetries discussed above,
%It has been demonstated that
 the one-dimensional Hubbard model possesses an infinite number of conservation laws~\cite{Shastry.86},
and is exactly solvable by Bethe-Anzatz~\cite{Ogata.1990,Schlottmann.1997,Guan.2013}. % which makes it integrable irrespective of the stength of the interaction.
Therefore, it is an excellent candidate to explore the link between  KPZ scaling, integrability and 
global non-abelian symmetries. 
%in  the charge and spin sectors, since it is integrable and (at half-filling) also it exhibits a global symmetry $SU_c(2)\otimes SU_s(2)$.}

\section{Quench Protocol}\label{sec:quench}
In the following, we present the scenario in which the initial state is prepared with an imbalance, either in the occupation or the magnetization.
In the infinite temperature state, the thermal energy of the particles is so high that there is no correlation or coherence among the particles. Therefore, the $T=\infty$ state is described by the diagonal density matrix
\begin{equation}
 \rho(T=\infty)=4^{-L}\prod_x \mathds{1}_x
\end{equation}
in which all local states are equally populated, i.e.,
\begin{equation}
 \mathds{1}_x =\ket{0}\bra{0} +\ket{\uparrow}\bra{\uparrow} +\ket{\downarrow}\bra{\downarrow} +\ket{\uparrow\downarrow}\bra{\uparrow\downarrow}.
\end{equation}
 
 We consider a  quench protocol  in which the system is prepared in an inhomogeneous state, 
 corresponding to a small deviation from the homogeneous $T=\infty$ state such that the initial
 state consists of two halves with a charge imbalance,
 \begin{equation}
\rho(t=0)=\prod_{x=-L/2}^{L/2}\rho_x(0)
\label{eq:rho_0}
 \end{equation}
 with
  \begin{equation}
  \rho_x(0) =\left \lbrace
    \begin{array}{cc}
        {1\over 4} \mathds{1}_x -{\mu\over 2}\ket{0}\bra{0}+{\mu\over 2}\ket{\uparrow\downarrow}\bra{\uparrow\downarrow}\;,\phantom{aa} x < 0 \\\\
   {1\over 4} \mathds{1}_x -{\mu\over 2}\ket{0}\bra{0}-{\mu\over 2}\ket{\uparrow\downarrow}\bra{\uparrow\downarrow}\;,\phantom{aa}  x\ge 0
    \end{array}
    \right. ,
    \label{eq:rho_0_n}
 \end{equation}
where $\mu \ll 1$ is a parameter that controls the deviation from the $T=\infty$ state, that corresponds to $\mu=0$.
The average occupation at any time $t$ can be expressed in terms of the density matrix as
\begin{equation}
  \average{n_x}(t)={\rm tr}\{\rho(t)\, c^\dagger_{x\sigma} c_{x\sigma}\}\label{eq:n_x}
\end{equation}
and evaluates  at $t=0$ to %can be evaluated analytically at $t=0$
\begin{equation}
 \average{n_x} (t=0) = \left \{
  \begin{array}{cc}
   1+{\mu\over 2}\;,\phantom{aaa} x<0 \\\\
   1-{\mu\over 2}\;, \phantom{aaa} x\ge 0
  \end{array}
 \right .
\end{equation}
At $t=0$, the density profile corresponds to a step function with a small imbalance in the occupations $\propto \mu$ between the two halves of the chain.
Notice that this initial state breaks the $SU_c(2)$ symmetry, but still preserves the spin $SU_s(2)$ symmetry.

As the initial state does not commute with the Hamiltonian~\eqref{eq:Hubbard}, the system  is
out of equilibrium, and undergoes non-trivial dynamics.
Numerically, we solve the von-Neuman equation that governs the evolution of the system's density matrix
\begin{equation}
  i\frac{\partial}{\partial t}\rho(t)=[H,\rho(t)],\label{eq:vN} 
\end{equation}
We solve Eq.~\eqref{eq:vN} by using the vectorization  procedure~\cite{amshallem2015approaches,Dzhioev.2011,Jiang.2013} $\rho(t)\to \ketL{\rho}$ within the matrix product state
framework~\cite{White.2004,Vestraete.2008,Moca.2022}. For that, we use the superfermion representation~\cite{Harbola.2008,Dzhioev.2011,Dzhioev.2012} which introduces a new set of
annihilation (creation) operators $\tilde c_{x,\sigma}^{(\dagger)}$ that satisfy the usual anticommuting relations, and generate
the dual Fock space~\cite{Dzhioev.2011}.

To obtain the time evolution of $\ketL{\rho(t)}$, we utilized the  %non-Abelian
 time-evolving block decimation ~\cite{Werner.2020,Vidal.2003, Vidal.2004, Verstraete.2004,Vestraete.2008},  
 %(NA-TEBD)
with abelian  symmetry operators, $U_c(1)\to \mathbb{U}_c(1)$ and $U_s(1)\to\mathbb{U}_s(1)$
extended to the vectorized  Liouville space~\cite{Moca.2022}. In addition, we also solve the problem  analytically  for the non-interacting system ($U=0$)  to benchmark our  TEBD computations.

Unlike usual TEBD simulations, which break down after relatively short times, here 
we can use TEBD simulations  with a small maximum bond dimension  up to rather long times and for 
relatively large system sizes  due to the slow growth of 
operator entanglement (entanglement of vectorized density matrix)~\cite{Ljubotina.2017}.
This is probably a consequence of having  initial states close to the $T=\infty$ state. 
%This guarantees that long-time evolutions can be performed with relative ease.
Computational time is further reduced by exploiting  abelian  charge and spin  symmetries 
 in the calculations. If not otherwise specified, results  presented in this work have been obtained
 by using a bond dimension for the multiplets $M=100$, 
 which guarantees an error of $\varepsilon < 10^{-7}$ in the discarded Schmidt coefficients.
 %, depending on the values of $U$. 
 Further details on the actual numerical method have been presented in Ref.~\cite{Moca.2022}.

%So far we have discussed just the charge transport, but
%it is important to note that these findings are applicable to any conserved quantity which can be associated with the
%existence of a non-Abelian symmetry. As discussed before, in particular, at half filling, the Hubbard model,
%has a $SU_s(2)$ spin symmetry, which allows us to test the spin-charge duality and construct the universal functions in the spin channel as well.
An analogous quench protocol can be implemented in the spin sector, which, in particular at half-filling, allows to test the spin-charge duality.
For that, we impose a weak imbalance in the magnetization at $t=0$.
The initial density matrix is given by~\eqref{eq:rho_0} with the local density matrices $\rho_x(0)$ given by
\begin{equation}
  \rho_x(0) =\left \{
    \begin{array}{cc}
         {1\over 4} \mathds{1}_x \gzb{+} {\mu_z\over 2}\ket{\uparrow}\bra{\uparrow}\gzb{-}{\mu_z\over 2}\ket{\downarrow}\bra{\downarrow} ,\phantom{aa} x < 0 \\\\
         {1\over 4} \mathds{1}_x\gzb{-}{\mu_z\over 2}\ket{\uparrow}\bra{\uparrow}\gzb{+}{\mu_z\over 2}\ket{\downarrow}\bra{\downarrow} , \phantom{aa} x\ge 0\\
    \end{array}
    \right. ,
    \label{eq:rho_0_m}
 \end{equation}
where $\mu_z$ controls spin polarization on the two halves of the chain, $\delta s_z(x,0)=\langle \frac 1 2 \sigma^z(x,t)\rangle_{t=0} = \pm \mu_z /2$
for $x<0$ and $x\ge 0$, respectively.

\begin{figure}[t!]
 \begin{center}
  \includegraphics[width=0.9\columnwidth]{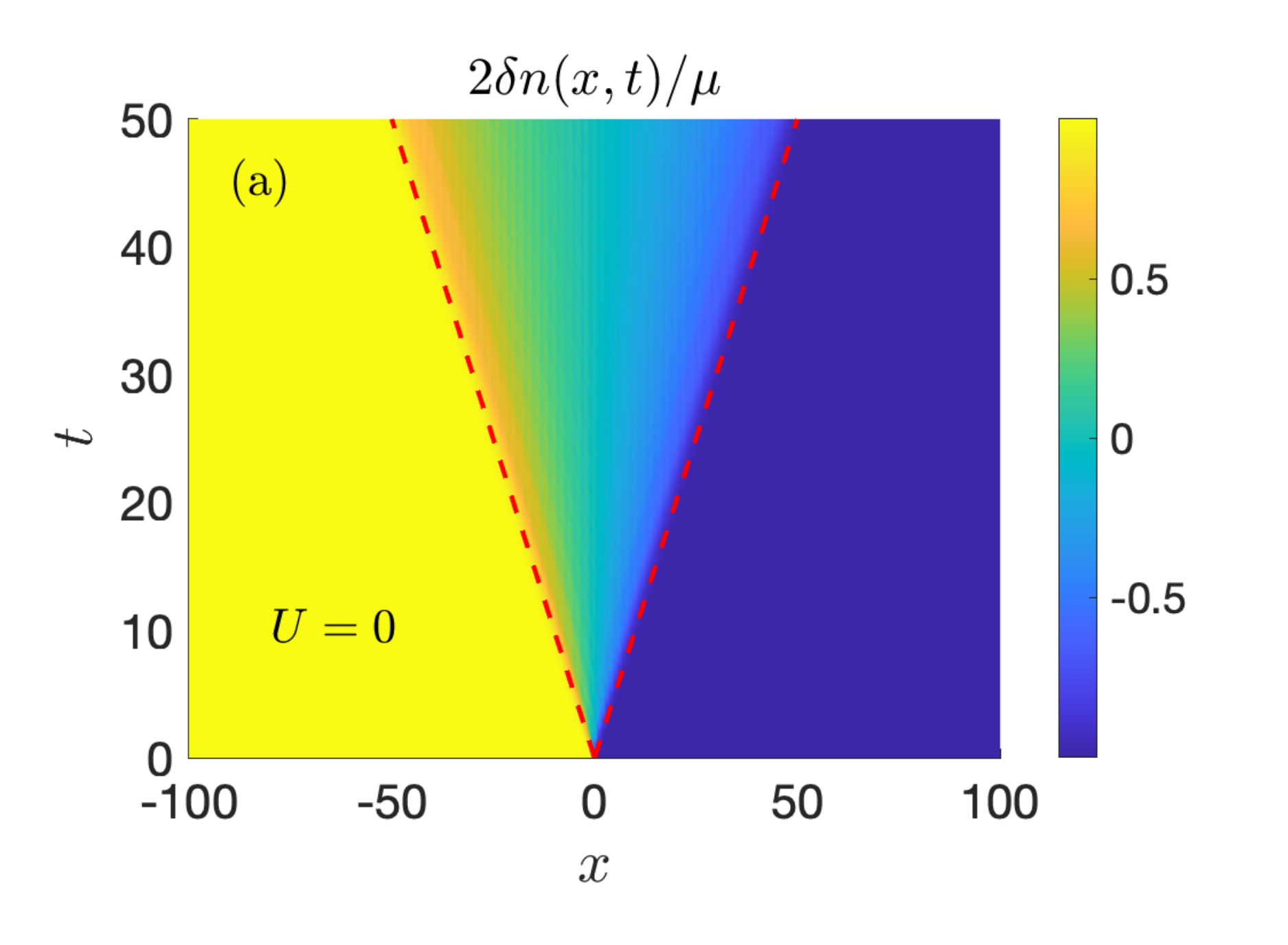}
  \includegraphics[width=0.9\columnwidth]{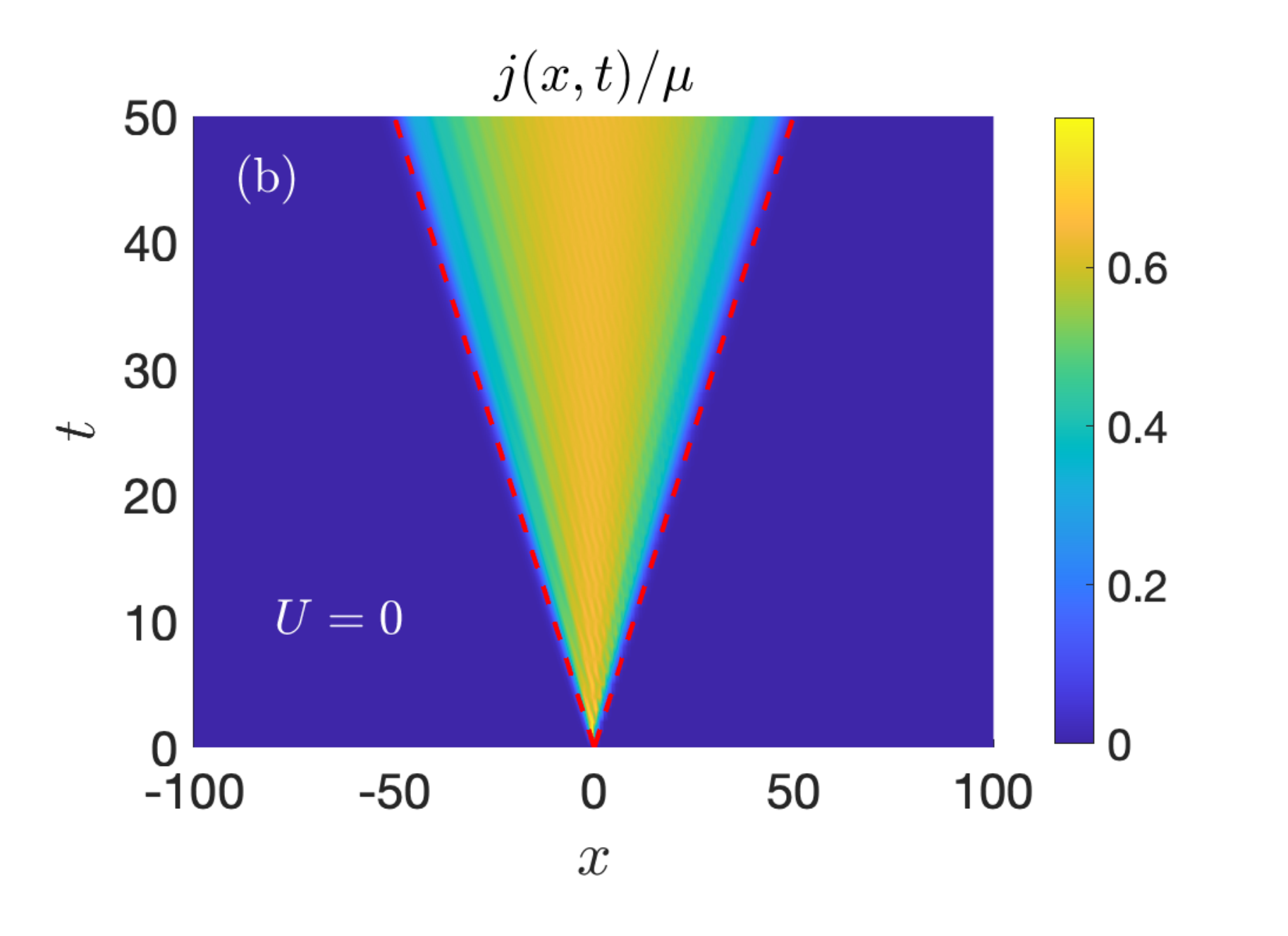}
  \caption{Heatmap of occupation density and charge current in the non-interacting case $U=0$.
  (a) Time evolution of the average occupation along the chain with respect to half filling $\delta n (x,t) = \average{n(x,t)}-\average{n}$, with $\average{n}=1$. 
  (b) Average current along the chain, demonstrating a regular light cone propagating ballistically with a constant Lieb-Robinson velocity of $v_F\approx J$. The red dashed lines provide a visual guide, corresponding to $t= x/v_F$. The system size is fixed at $L=200$ sites.}
  \label{fig:ballistic}
 \end{center}
\end{figure}
\begin{figure}[t!]
 \begin{center}
  \includegraphics[width=0.9\columnwidth]{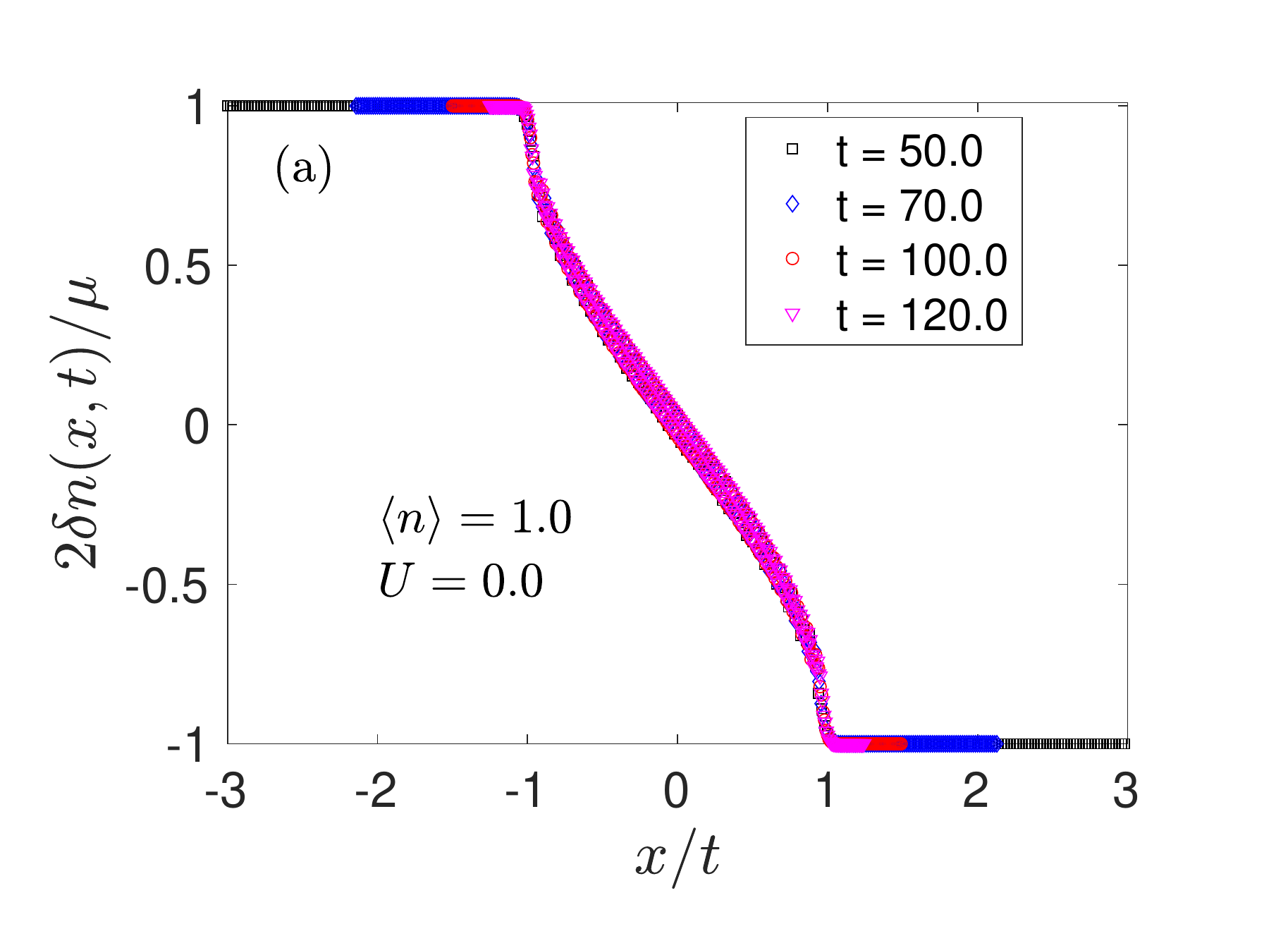}
  \includegraphics[width=0.9\columnwidth]{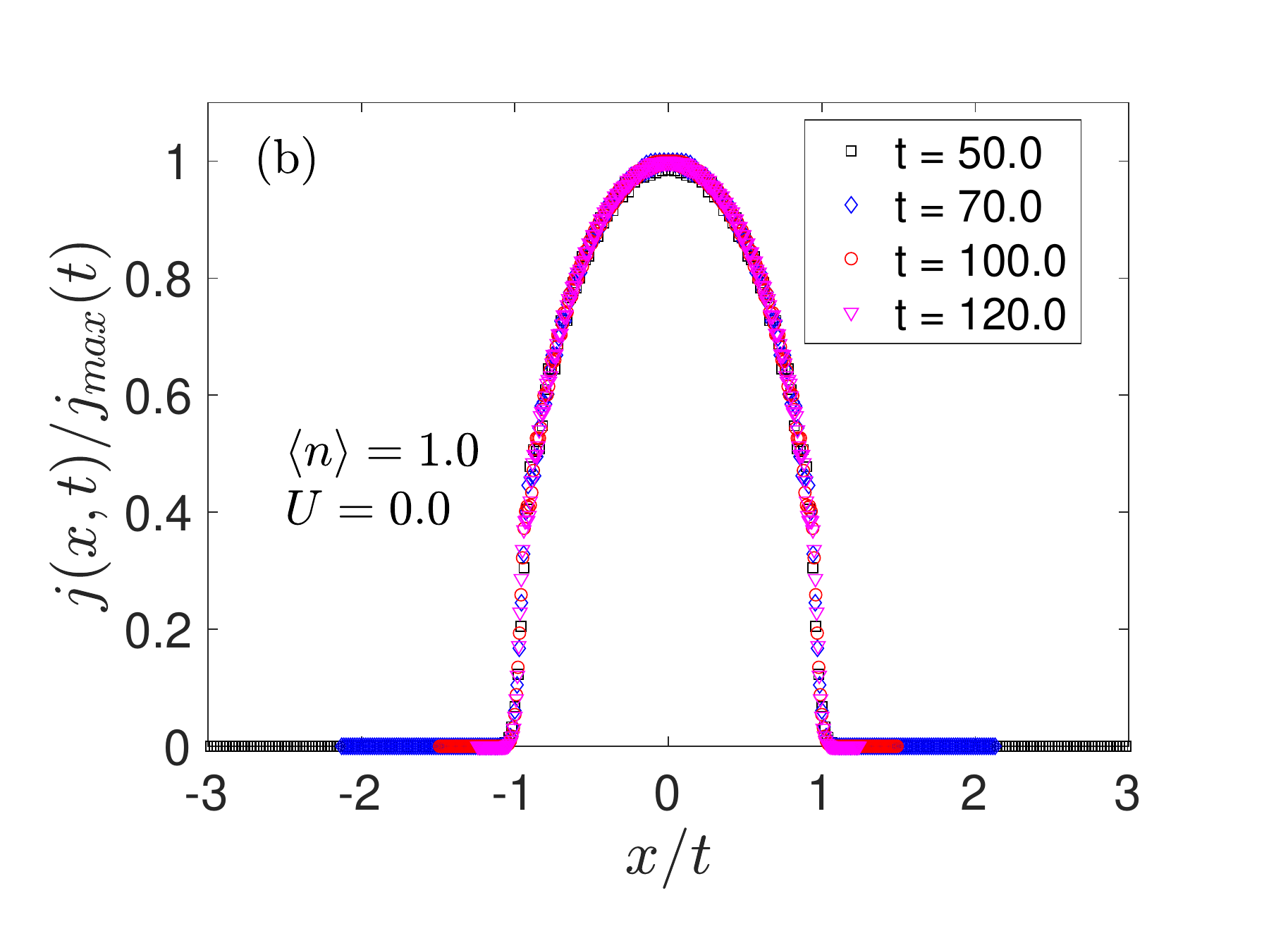}
  \caption{(a) Rescaled profiles for the average occupation $\delta n(x,t)$ at $U=0$ and half filling at various times, 
  plotted
  %G represented 
  as a function of $x/t$ to display the universal ballistic scaling.
  (b)The corresponding current profile, $j(x,t)$ 
  %along the chain at different times 
  exhibits a similar ballistic scaling. }
  \label{fig:scaling_ballistic}
 \end{center}
\end{figure}

\begin{figure}[t!]
 \begin{center}
  \includegraphics[width=0.9\columnwidth]{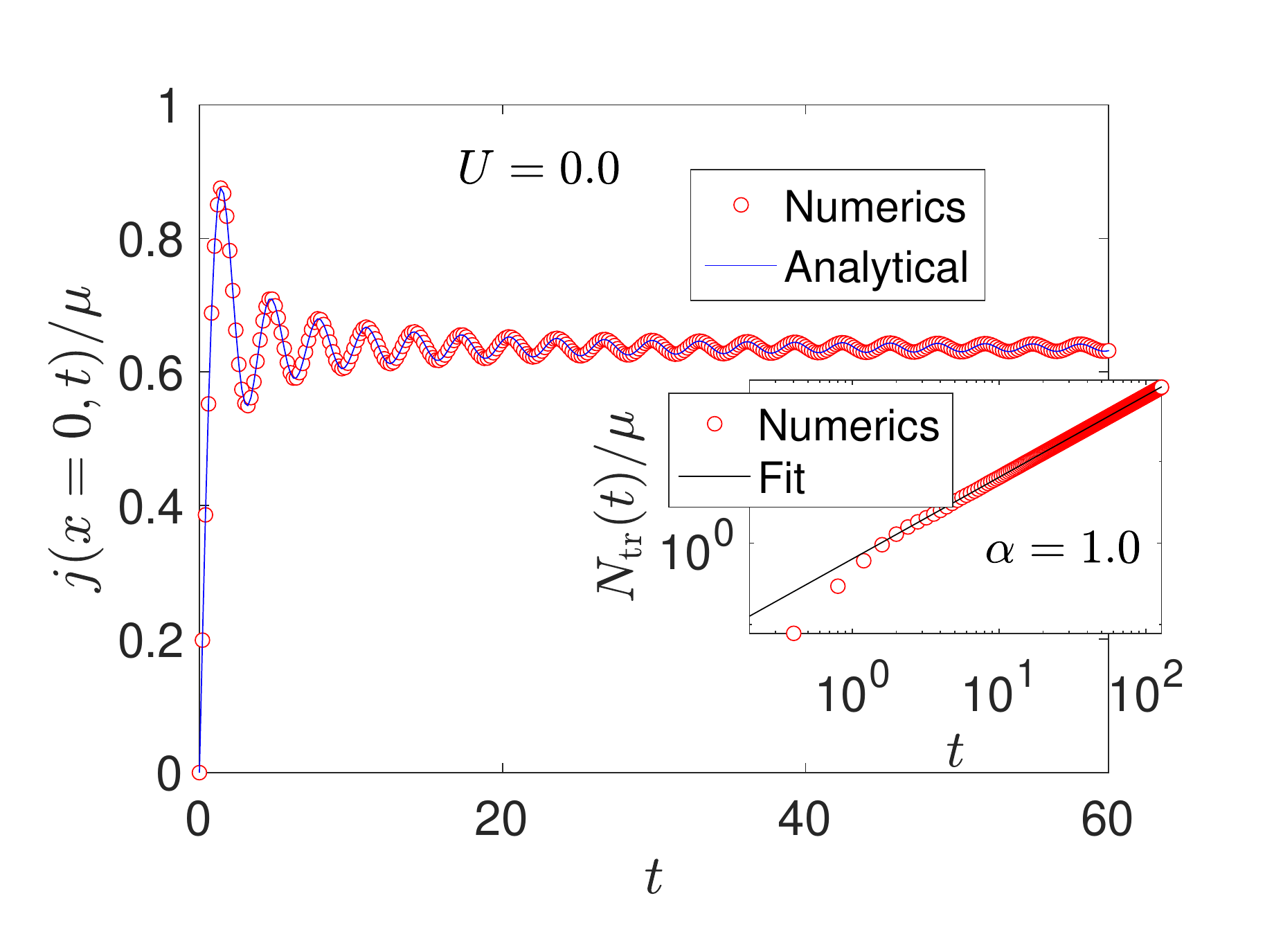}
  \caption{Time evolution of the current $j(x=0, t)$ across the interface at $U=0$.
  The symbols correspond to the numerical results obtained using TEBD, while the solid line represents the analytical expression \eqref{eq:current_analytical}.
  The transient oscillations decay as $\propto t^{-1}$, and the current remains constant in the limit of $t\to \infty$, indicating a ballistic behavior. The inset shows the total number of particles that tunnel
  across the interface $N_{\rm tr}(t)$ together with the power law fitting, which indicates a scaling exponent $\alpha\approx 1$ associated with ballistic transport. }
  \label{fig:current_across_ballistic}
 \end{center}
\end{figure}
\section{Ballistic behavior ($U=0$)}\label{sec:ballistic}
In the non-interacting limit,  $U=0$, the quasiparticles 
move independently and are not scattered by each other.
Their dispersion relation is given by the usual tight-binding formula, 
$\varepsilon_k=-J\cos(k)$, which determines their maximal quasiparticle  velocity, $v_\text{max}= J = 1$, 
 identified as the  Lieb-Robinson velocity~\cite{Cheneau2012,Gogolin_2016}.
%and they have a well-defined velocity.
In the absence of interactions, quasiparticles motion is coherent, 
and transport is referred to as 'ballistic', implying a linear relation between  distance   and traveling time.
While in this section we present results specific 
to charge transport at half filling, it is important to remark that for $U=0$ ballistic transport persists  
for any filling,
%$\average{n}>0$ in the non-interacting limit ($U=0$), in 
both in the charge and in the spin sector.

In Fig.~\ref{fig:ballistic} we display the evolution of the average occupation $\delta n (x,t) = \average{n(x)}(t)-\average{n}$ and the particle current 
\begin{equation}
  j(x,t) = {\rm tr}\Big\{{i\over 2} \big[c^{\dagger}_{x+1,\sigma} c_{x\sigma}-c^\dagger_{x\sigma} c_{x+1\sigma}\big]\rho(t)\Big\},
  \label{eq:current}
\end{equation}
along the chain after the initial quench.
%It indicates 
These both display  light-cone propagation  of quasiparticles with a constant velocity, $v_\text{max}$.
%whose maximal possible value is
% the Lieb-Robinson velocity~\cite{Cheneau2012,Gogolin_2016}
%associated with the Fermi velocity $v_F\approx J$.
%
%G there is no Fermi velocity...
%
To %determine the specific type of 
confirm ballistic transport, we performed quantitative analysis of charge and current equilibration, 
%This includes studying t
and the scaling of charge and current profiles.
%A clear indication of ballistic scaling can be observed in 
Fig.~\ref{fig:scaling_ballistic} demonstrates ballistic scaling for $U=0$: %where
 the rescaled profiles for $\delta n(x,t)$ and $j(x,t)$ collapse onto  a single universal curve, 
 %demostrating a universal behavior  as a function of
when plotted against   $x/t$.
The total number of particles transferred across the interface %, between the two halves of the chain 
is another useful quantity, whose asymptotic behavior in the long time limit allows to identify the type of dynamics~\cite{Marko.2019, Ye.2022}. 
%The asymptotic scaling exponent $\alpha=1/z$, which characterizes the universalty class, is determined from 
The total charge across the interface scales as
\begin{equation}
  N_{\rm tr}(t) = \int_{0}^t j(0, t')dt'\propto t^\alpha, 
\end{equation}
with an exponent $\alpha=1/z$.  For ballistic transport, one has $\alpha=1$, while diffusive transport is characterized 
by $\alpha=1/2$, and anomalous diffusion by an exponent different from these.
As Fig.~\ref{fig:current_across_ballistic} shows, $j(0,t)$ exhibits a rich structure. 
Following the quench, the current displays transient oscillations with an approximate frequency $\omega\approx \omega_J\equiv J$, and  
an amplitude decaying as $\propto t^{-1}$. In the long-time limit, the current approaches a finite asymptotic value. %and does not decay.
The inset displays the total number of particles transferred across the interface,  increasing linearly with time, $N_{\rm tr} (t)\propto t$, 
corresponding to
% an exponent $\alpha \approx \alpha_b = 1$, associated with 
ballistic transport.

Our TEBD results are consistent with the analytical findings. 
By assuming periodic boundary conditions (PBC) and performing a Fourier transform of the time evolution,
 it is possible to express the time dependence of the annihilation (creation) operators in the Heisenberg picture in terms of  Bessel functions of the first kind,
\begin{equation}
c_{\tilde x\sigma}(t)=\sum_{x}i^{x-\tilde x} J_{x-1}(\omega_J\, t)c_{\tilde x, \sigma} ,
\end{equation}
with the frequency associated with the hopping integral, $\omega_J=J$.

\begin{figure}[tb]
 \begin{center}
  \includegraphics[width=0.9\columnwidth]{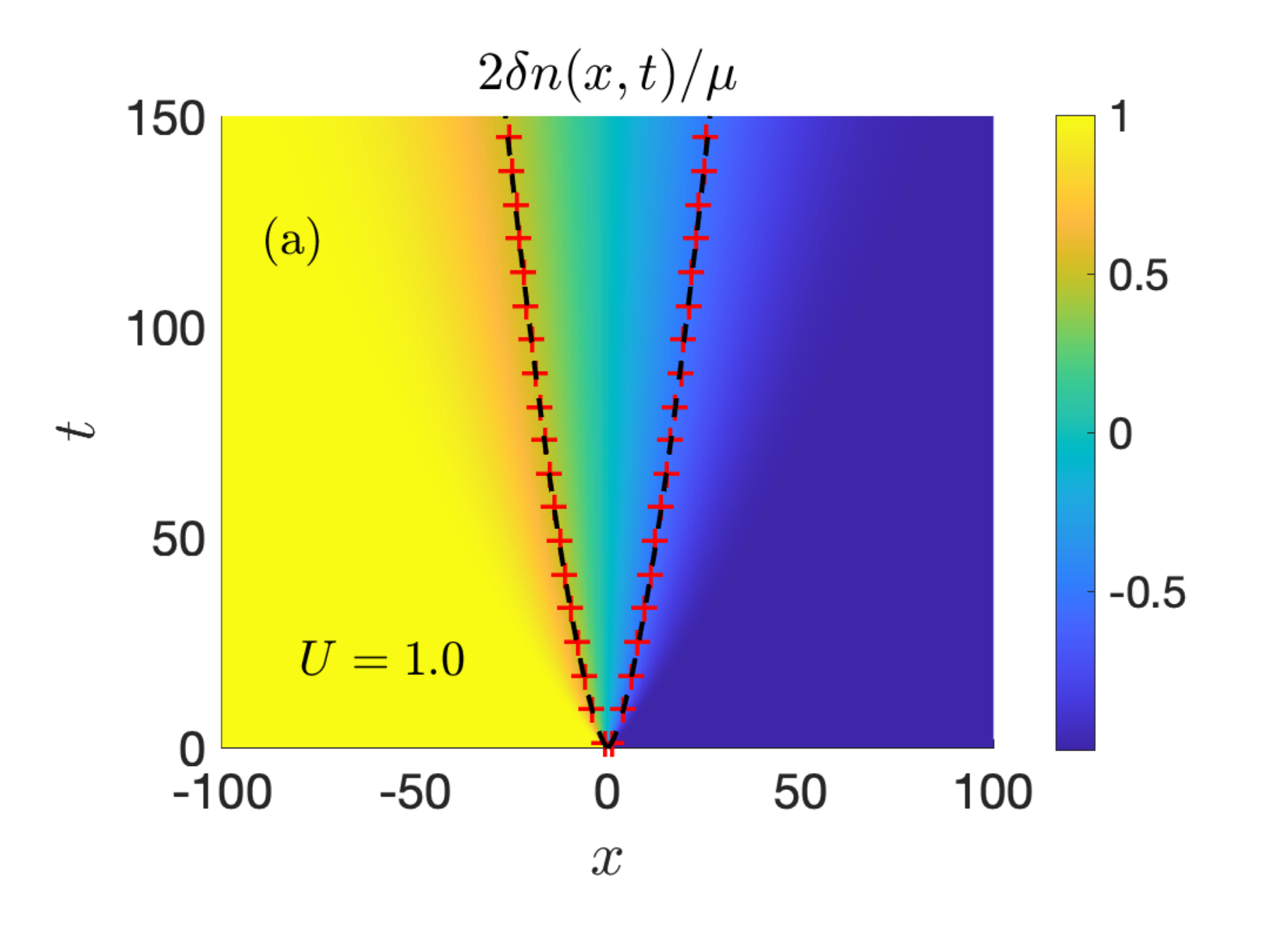}
  \includegraphics[width=0.9\columnwidth]{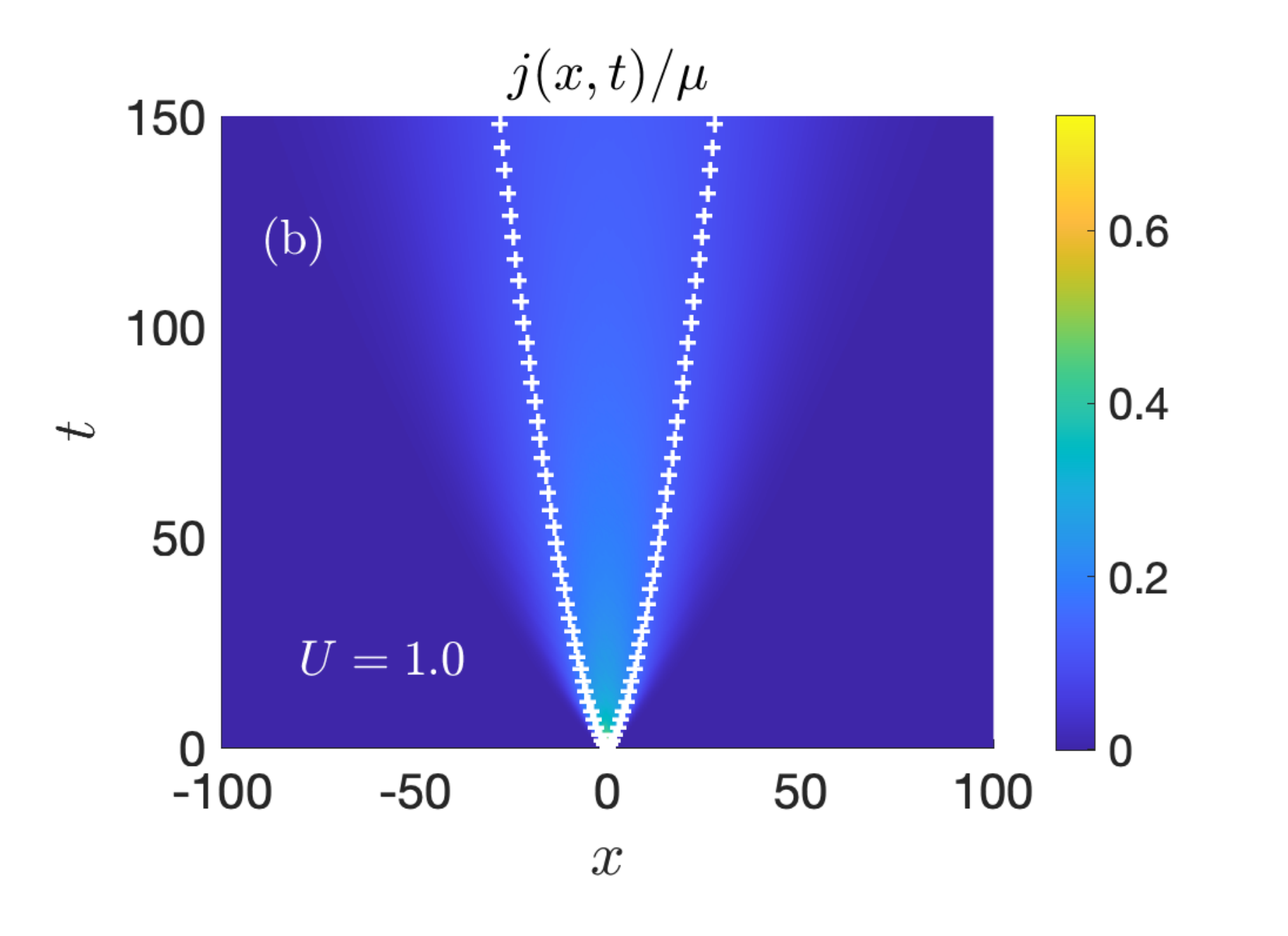}
  \caption{Spatio-temporal evolution of the average occupation $\delta n(x,t)$ and the average current $\langle j(x,t)\rangle$ along the chain at half-filling with a finite interaction strength $U=1.0$, indicate a superdiffusive behavior. Symbols correspond to  mid-value contours, where  $2\delta n(x,t)/\mu = \gzb{\pm}0.5$,
    while dashed lines indicate $t^{2/3}=r\,x/v_F$, with a scaling factor $r=1.1$. }
  \label{fig:superdiffusive}
 \end{center}
\end{figure}
By solving the von Neumann equation~\eqref{eq:vN} for the density matrix $\rho(t)$, we can determine the 
%average 
expectation value of the current operator at site $x$ along the chain. Using the standard approach of calculating the trace of the product of the density matrix and the current operator, Eq.~\eqref{eq:current}, we obtain the current at the interface $x=0$ at any later time,
\begin{equation}
j(0,t) = \mu \sum_{x=-L/2}^{L/2-1}\sum_{\sigma} J_x(\omega_J t)\,J_{x+1}(\omega_J t).
\label{eq:current_analytical}
\end{equation}
This  result  matches with the TEBD data, as shown in the main panel of Fig. \ref{fig:current_across_ballistic}.

\section{Superdiffusive behavior at half-filling ($U\ne 0$, $\average{n}=1$)}  \label{sec:superdiffusive}
\begin{figure}[tbh!]
 \begin{center}
  \includegraphics[width=0.9\columnwidth]{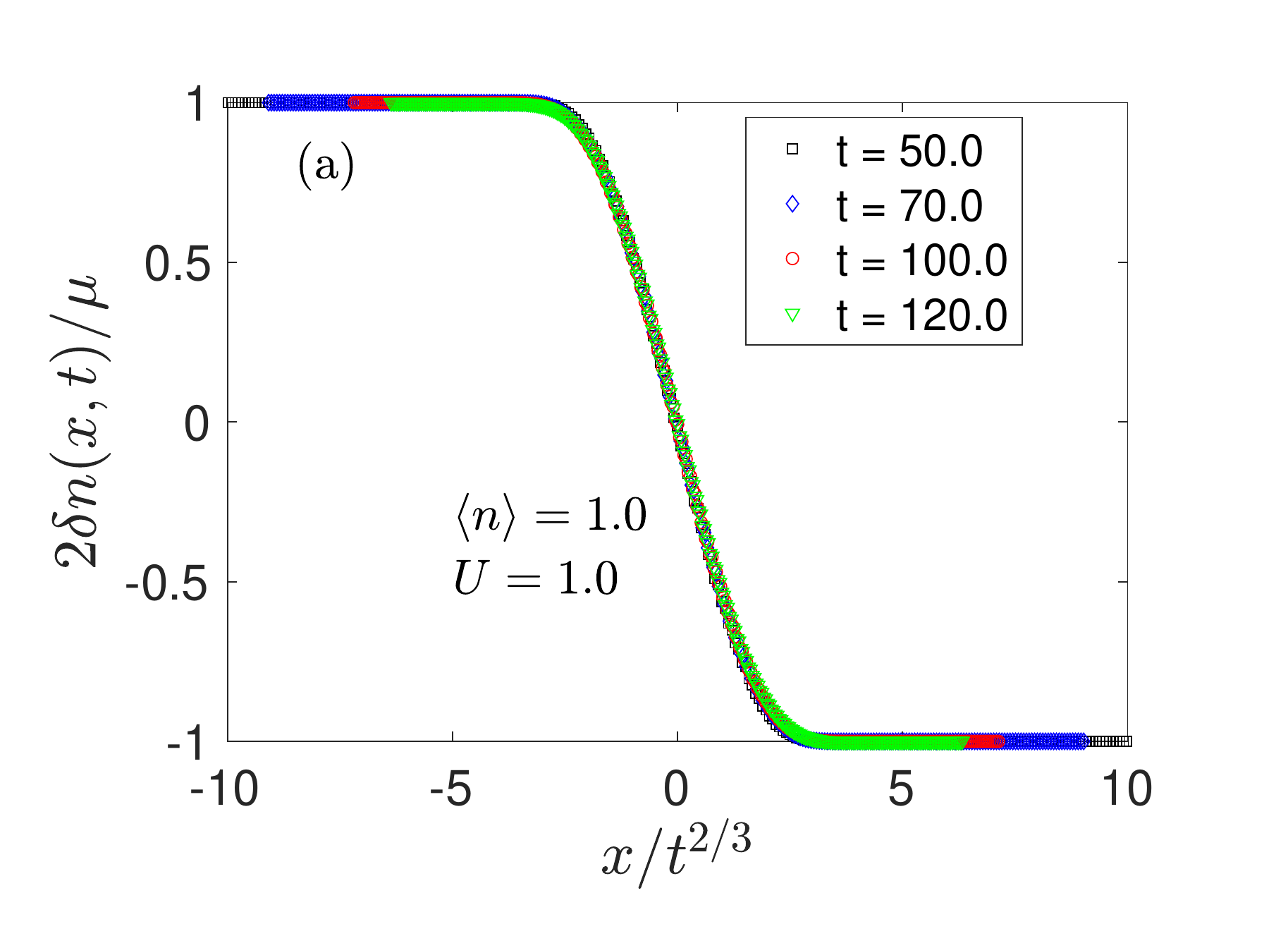}
  \includegraphics[width=0.9\columnwidth]{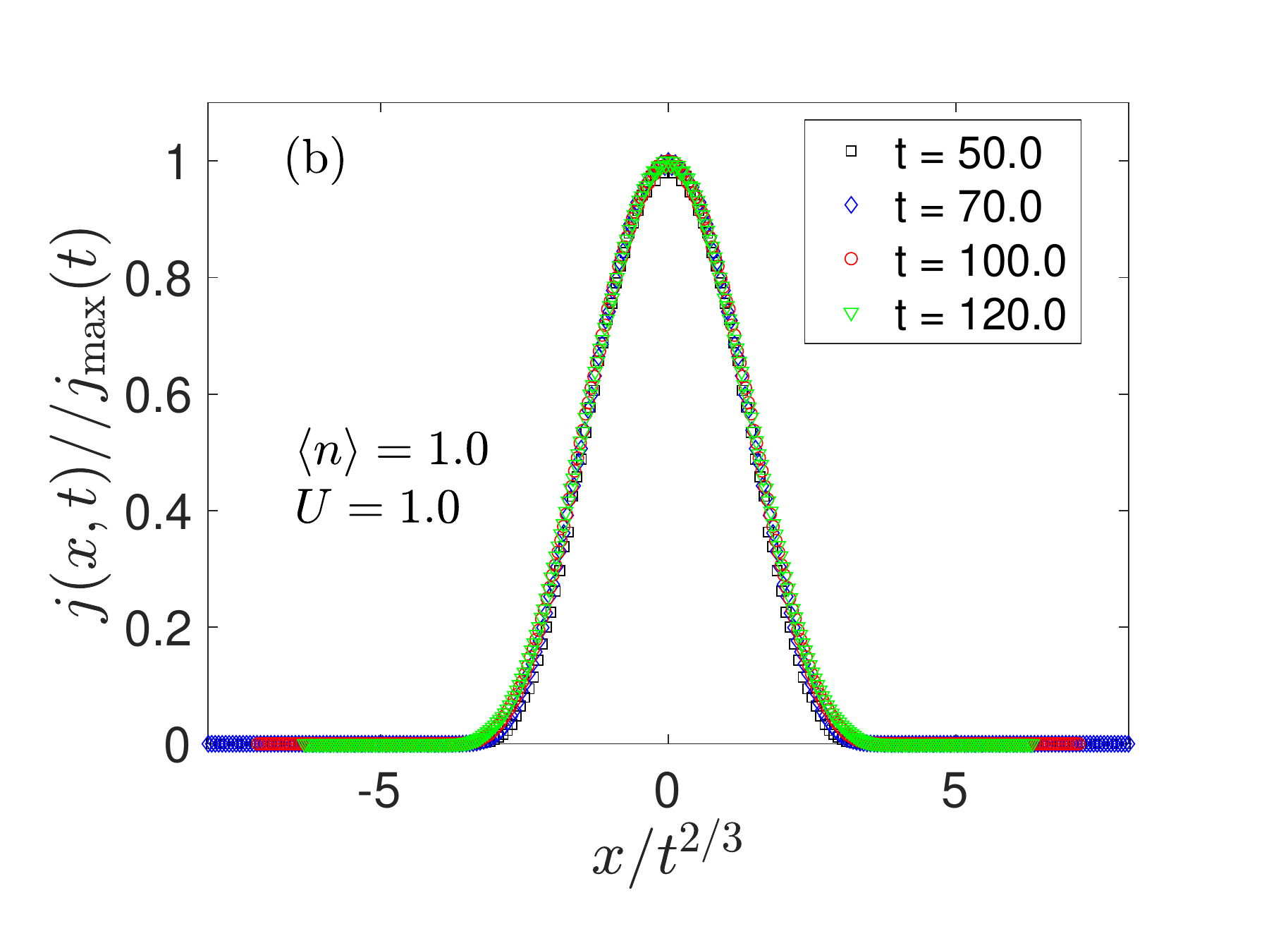}
  \caption{Rescaled profiles of (a) the average occupation $\delta n(x,t)$, and (b) the current $j(x,t)$, at half filling for $U=1.0$
   at various times. 
       Both display universal superdiffusive scaling, in terms of $x/t^{2/3}$.
       %$\delta n(x,t) \propto \gzb{\rho(b\,x/t^{2/3})}$.
       }
  \label{fig:scaling_superdiffusive_n_1}
 \end{center}
\end{figure}

In general, superdiffusive behavior refers to a %type of 
diffusion process in which the mean square displacement of quasi-particles increases faster than linear with time, but slower than quadratic.
This behavior is often characterized by  an asymptotic power-law relationship 
\be
\langle \Delta x^2\rangle \approx 2\,D \; t^{2\alpha}, 
\ee
with an exponent \tp{$\alpha$} greater than  1/2  and smaller than 1, and $D$,  the anomalous  
 diffusion constant. In the specific case of the KPZ scaling, 
the superdiffusive exponent is $\alpha_\text{sd}=2/3$.

In this section, we study transport in the presence of interactions at half filling, $\langle n\rangle =1$, 
and demonstrate superdiffusive behavior.
%The model exhibits two non-abelian $SU(2)$ symmetries related to the total 
At half-filling, pseudo-charge and  spin $SU(2)$  symmetries are related by duality, which implies that 
spin and charge transport  have identical properties.
%  that the transport coefficients 
%of the model are described equally well by either spin or charge degrees of freedom. 

\begin{figure}[t!]
 \begin{center}
  \includegraphics[width=0.9\columnwidth]{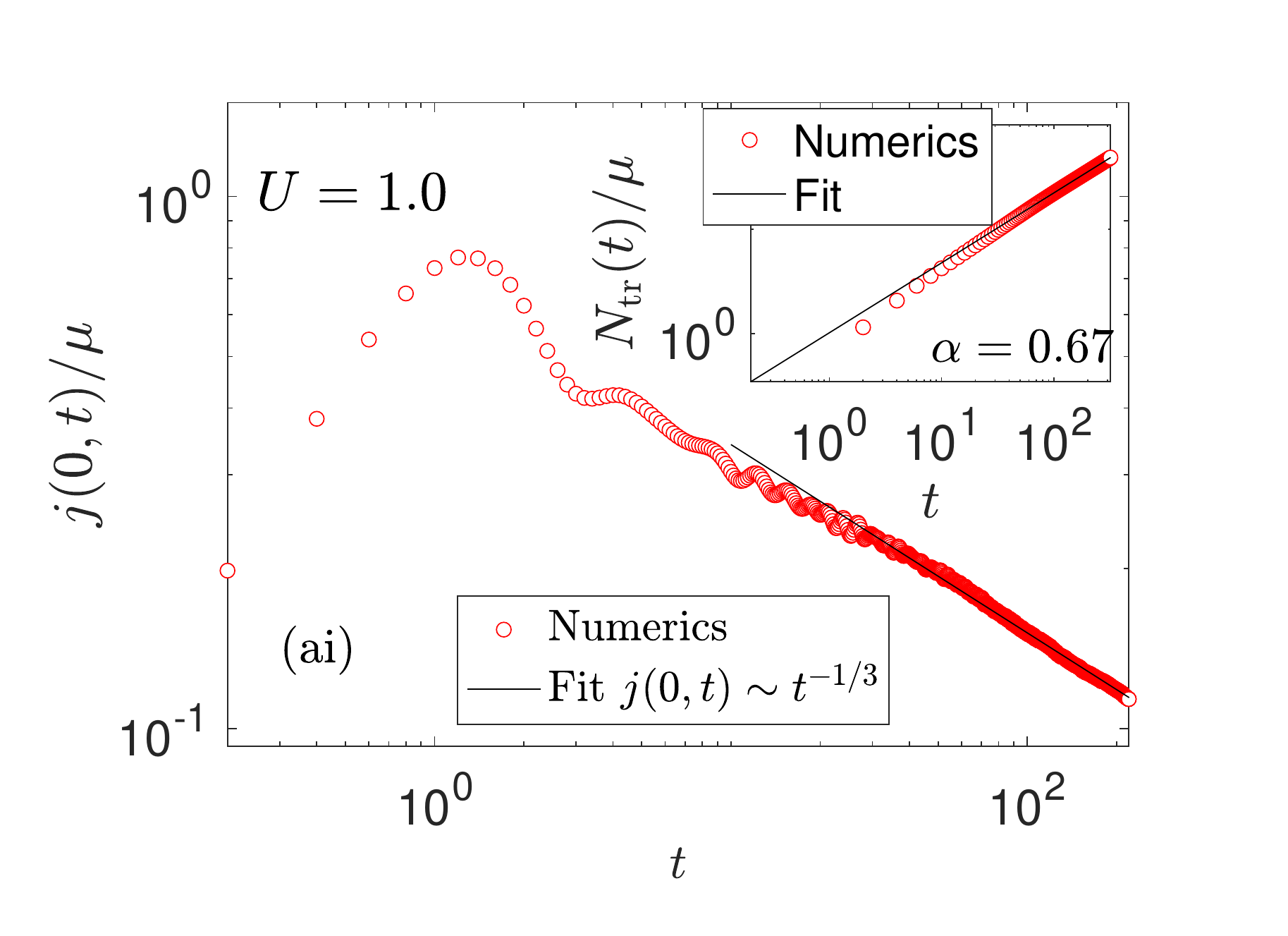}
  \caption{
  %(a) 
  The time dependence of the current across the interface follows a power-law decay, 
  $\sim t^{1-\alpha}$ with a superdiffusion exponent, $\alpha_{sd}\approx 2/3$. 
  The transferred charge scales as $N_\text{tr}(t)\sim t^{\alpha} = t^{2/3}$, as
  shown in the inset. %, with the actual scaling displayed. 
  System size is $L=300$ sites.
%  (b) The charge diffusion coefficient $D_c$ \gzb{\bf we  use $D_c$ for the charge diffusion constant}
%  as a function of the interaction strength $U$. This relationship is shown in the main plot, while the inset displays the same data on a logarithmic scale. The dashed line in the inset corresponds to a behavior proportional to $D\sim 1/U$.
% \gzb{\bf I do not understand the normalization of the  charge diffusion constant.
% How is it compatible with $b=0.98$ in Fig. 7? One could determine it directly from the charge correlation function.} 
  }
  \label{fig:scaling_superdiffusive_n_2}
 \end{center}
\end{figure}

\subsection{Charge sector}
With the initial density matrix $\rho(0)$ as defined in Eq.~\eqref{eq:rho_0}, we assess the transport properties in the charge sector by analyzing the decay of the average density and current profiles.
In Fig.~\ref{fig:superdiffusive} we display the spatio-temporal evolution of $\delta n(x,t)$
and the average current, $j(x,t)$, which both indicate slower than ballistic propagation of the quasiparticles.

%By following the same methodology as described in Sec.~\ref{sec:ballistic}, we demonstrate that the d
Indeed, the density and the current density profiles can be both presented as a function of a single scaling variable, $x/t^{\alpha}$, 
and collapse onto a single curve at large times, with $\alpha_\text{sd} = 2/3$, as predicted by
KPZ scaling. This is illustrated in Fig.~\ref{fig:scaling_superdiffusive_n_1}(a,b), where $\delta n(x,t)$ and $j(x, t)$ 
are  plotted against $x/t^\alpha$ for $U=1$, and shown to exhibit clear superdiffusive scaling with   $\alpha= 2/3$.

These results are corroborated with Fig.~\ref{fig:scaling_superdiffusive_n_2}, showing the power law decay of the current across the interface
in the long time limit,  $j(0, t)\sim t^{\alpha -1} = t^{-1/3}$, implying  that the total number of particles transferred across the interface
scales as, $N_{\rm tr}(t)\sim t^{\alpha} =t^{2/3}$  (see inset).

\begin{figure}[t!]
 \begin{center}
  \includegraphics[width=0.9\columnwidth]{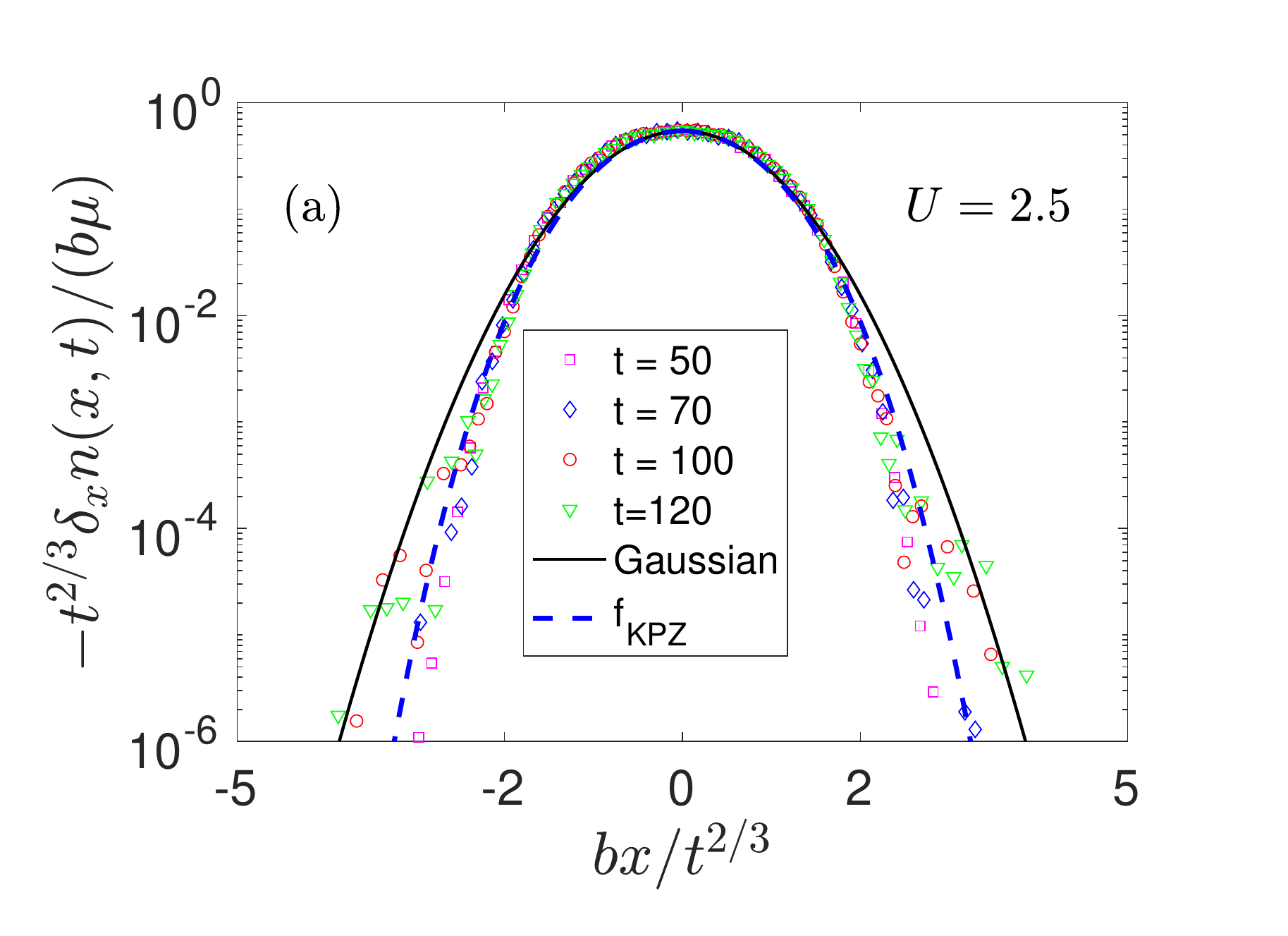}
  \includegraphics[width=0.9\columnwidth]{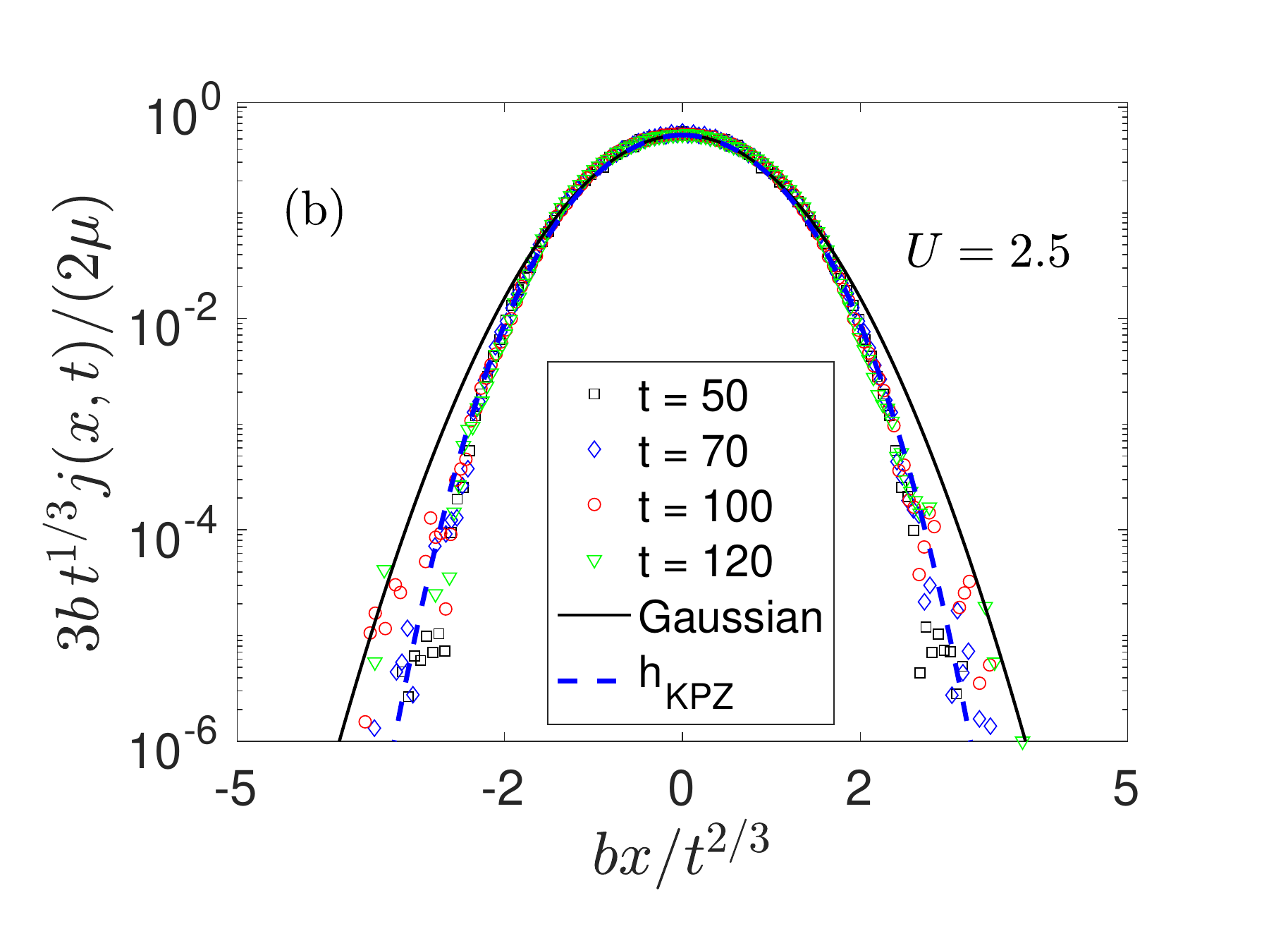}
  \caption{ The density gradient  $\delta_x n(x,t) = \delta n(x,t) - \delta n(x-1,t)$ (panel a), 
  as well as the  current density $j(x,t)$ (panel b) display universal KPZ scaling.
  Here $f_\text{KPZ}$ is the KPZ scaling function~\cite{Prahofer2004}, 
while $h_\text{KPZ}$ has been computed by integrating Eq.~\eqref{eq:h'}. Notice that 
a single scaling parameter $b=0.98$  is used to fit all profiles simultaneously
on both panels.
%\gz{{\bf Please, change axis label}
%I think the right scaling is: $3\,b\,j(x,t) \,t^{1/3}/2/\mu  = h(b x/t^\alpha)$, 
%and $- \delta_x n(x,t) \,t^{2/3}/b/\mu  = f(b x/t^\alpha)$. }
}
  \label{fig:scaling_superdiffusive}
 \end{center}
\end{figure}

 The current $j$ and  density gradient $\partial_x n$ exhibit similar, universal profiles~\cite{Marko.2019}.
The density profile, in particular, exhibits 
a scaling, 
\be
\delta n (x,t) =  \mu \, \rho\bigl( \,{b\,x} / {t^\alpha})\label{eq:n_scaling}
\ee
with $\rho (y)$ a  scaling function satisfying  $\rho (\mp\infty) = \pm 1/2$, 
and the scaling factor, $b$,  related to the diffusion constant,  $b\sim 1/\sqrt{D}$.
This immediately leads to the density gradient scaling as,
\be
\partial_x n (x,t) =- \frac {b\,\mu} {t^{\alpha}} \,f\bigl( \,{b\,x} / {t^\alpha})\;,
\label{eq:grad_n_scaling}
\ee
with the scaling function $f(y) = - \rho'(y)$. Notice that $f(y)$ obeys the sum rule,  
$\int \text{d}y\, f(y) =1$.

The scaling form of the current follows from the continuity equation, $\dot n (x,t) + \partial_x j(x,t) =0$, 
implying
\be
j (x,t) =   \frac {\mu}{b}  \,\alpha\, {t^{\alpha-1}} \,h\bigl( \,{b\,x} / {t^\alpha})\;,
\label{eq:j_scaling}
\ee
with the scaling function $h(y)$ related to $f(y)$ by the differential equation, 
\be 
h'(y) = - y\, f(y).
\label{eq:h'}
\ee
As shown in Fig.~\ref{fig:scaling_superdiffusive}, $j(x,t)$ and $\partial_x n(x,t)$ 
indeed satisfy the scaling forms, \eqref{eq:j_scaling} and \eqref {eq:grad_n_scaling}.
We emphasize that these scaling forms are just simple consequences of the rather natural scaling 
ansatz, \eqref{eq:n_scaling} and the conservation of electron charge.

We remark that in case of  a Gaussian current profile, $h(y)\sim e^{- \kappa\, y^2/2}$, $h(y)$ and $f(y)$
assume the same, Gaussian forms (apart from a prefactor). However, 
this relation is violated for any other profile, including the KPZ profile discussed here.

To  establish the system's universal dynamics conclusively, we now compare 
the scaling functions $f(y)$ and $h(y)$ to the KPZ scaling function.
For small $\mu$, one can use liner response theory to prove that 
the equilibrium correlation function, $\chi_n(x,t) \equiv \langle n(x,t) n(0,0) \rangle$, 
and the gradient $ \partial_x  n(x,t)$ are proportional to each other~\cite{Marko.2019},
\begin{eqnarray}
  \chi_n(x,t) & = & - \lim_{\mu\to 0  {1\over \mu} \partial_x \langle n(x,t)\rangle}
  \approx  -{1\over \mu} \delta_x  n(x,t)
\end{eqnarray}
where $\delta_x n(x,t) = \langle n(x,t)\rangle -\langle n(x-1,t)\rangle $ stands for the finite  
difference of the density.

The gradient profile in our set-up is therefore just the  density-density 
correlation function~\cite{Marko.2019}, which is well documented  in the KPZ model. 
The numerically obtained density gradient and current profile 
scaling functions  are displayed in Figs.~\ref{fig:scaling_superdiffusive}~(a,b), 
together with the tabulated KPZ ($f_\text{KPZ}$)  and simple gaussian  scaling functions~\cite{Prahofer2004},  
and $h_\text{KPZ}$ computed from Eq.~\eqref{eq:h'}.
Our numerical results show that the KPZ scaling functions accurately describe both the density gradient and the 
average current  profiles,  while the Gaussian scaling function does not. 
These results, together with the scaling properties of 
charge transfer and the deformation of the light cone, establish KPZ scaling of the charge correlations 
in the half-filled infinite temperature state.

\begin{figure}[t!]
 \begin{center}
  \includegraphics[width=0.85\columnwidth]{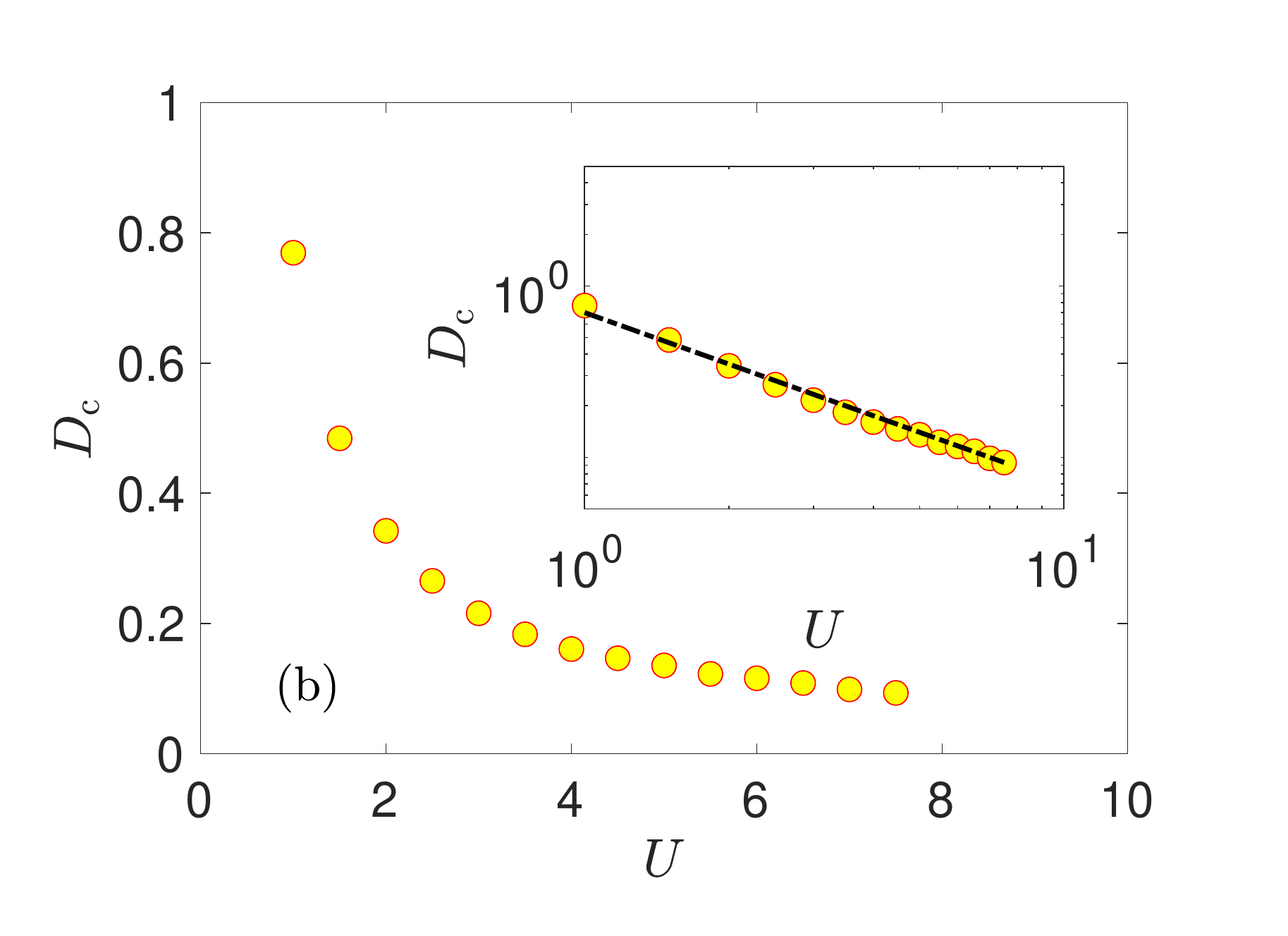}
  \caption{
%  (a) The time dependence of the current across the interface follows a power-law decay, 
%  \gzb{$\sim t^{1-\alpha}$ with a superdiffusion exponent, $\alpha_{sd}\approx 2/3$}. 
%  The transferred charge \gzb{scales as $N_\text{tr}(t)\sim t^{\alpha} = t^{2/3}$, as} 
%  shown in the inset.%, with the actual scaling displayed. 
%  System size is $L=300$ sites.
 % (b) 
  Charge diffusion coefficient $D_c$ as a function of the interaction strength $U$. This relationship is shown in the main plot, while the inset displays the same data on a logarithmic scale. The dashed line in the inset corresponds to a behavior proportional to $D_c\sim 1/U$.
  }
  \label{fig:diffusion_constant}
 \end{center}
\end{figure}

Let us close this section by discussing  the interaction dependence of the anomalous diffusion constant. 
Classically, the correlation function, $\chi_n(x,t) \equiv \langle n(x,t) n(0,0) \rangle$, is just the 
probability that a particle originally at  $x=0$ reaches position $x$ at time $t$, 
\be 
P(x,t|0,0) = \chi_n(x,t) \;.
\ee
The scaling form of the gradient,  \eqref{eq:grad_n_scaling} therefore implies, 
\be 
P(x,t|0,0) = \frac {b} {t^{\alpha}} \,f\bigl( \,{b\,x} / {t^\alpha})  \;.
\ee 
We can then express the variance of the distance as 
\be 
\langle x^2\rangle = \int \text{d}x\, x^2 \,P(x,t|0,0) = \,2\, D_c\, t^{2\alpha} \;,
\ee
with the charge diffusion constant expressed as 
\be 
 D_c = {\cal I } / {b^2}, 
\ee
with ${\cal I}=\frac1 2 \int \text{d}y\, y^2 \,f_\text{KPZ}(y) =0.25$ for the KPZ scaling function.

\begin{figure}[t!]
 \begin{center}
  \includegraphics[width=0.8\columnwidth]{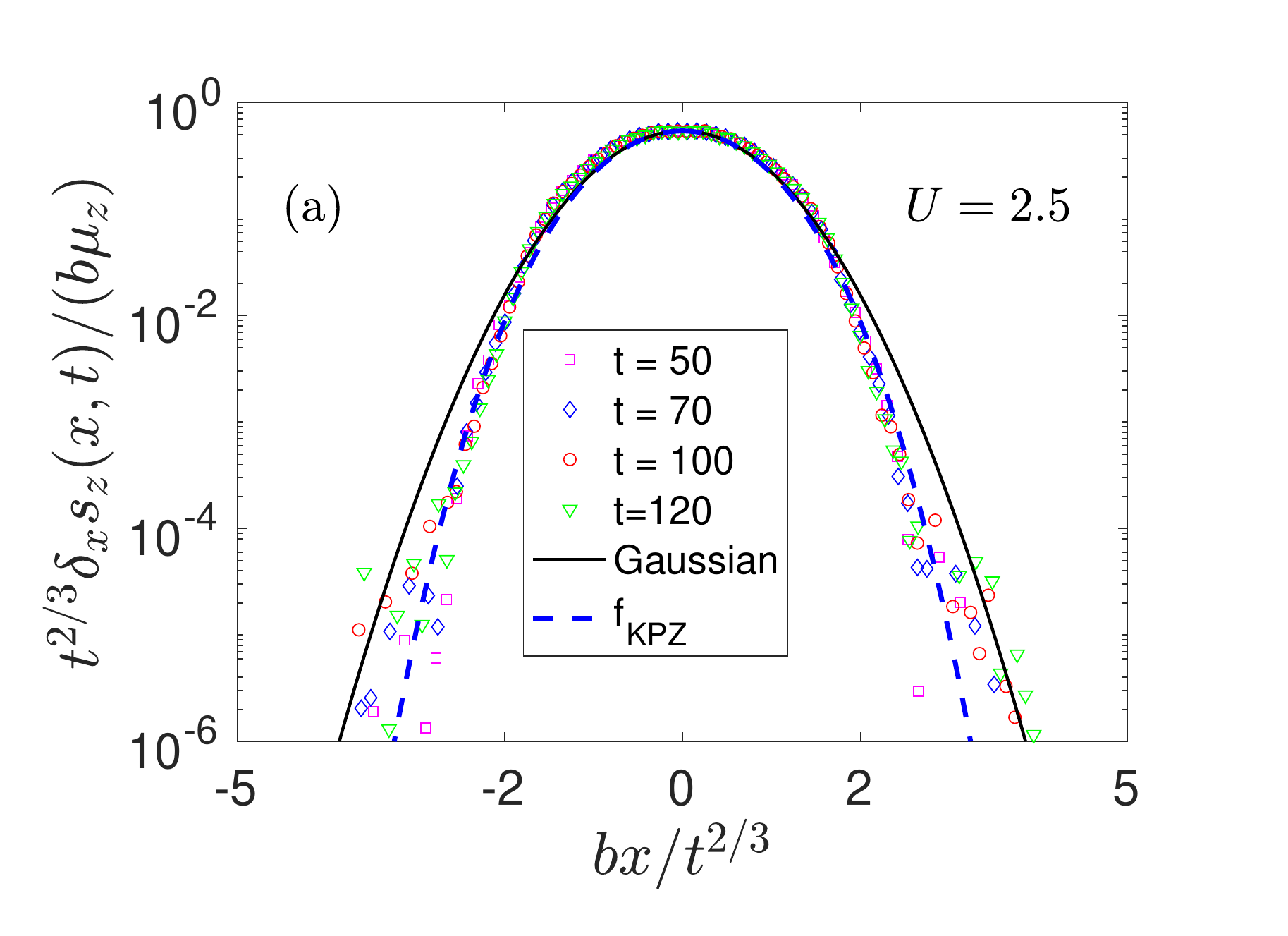}
  \includegraphics[width=0.8\columnwidth]{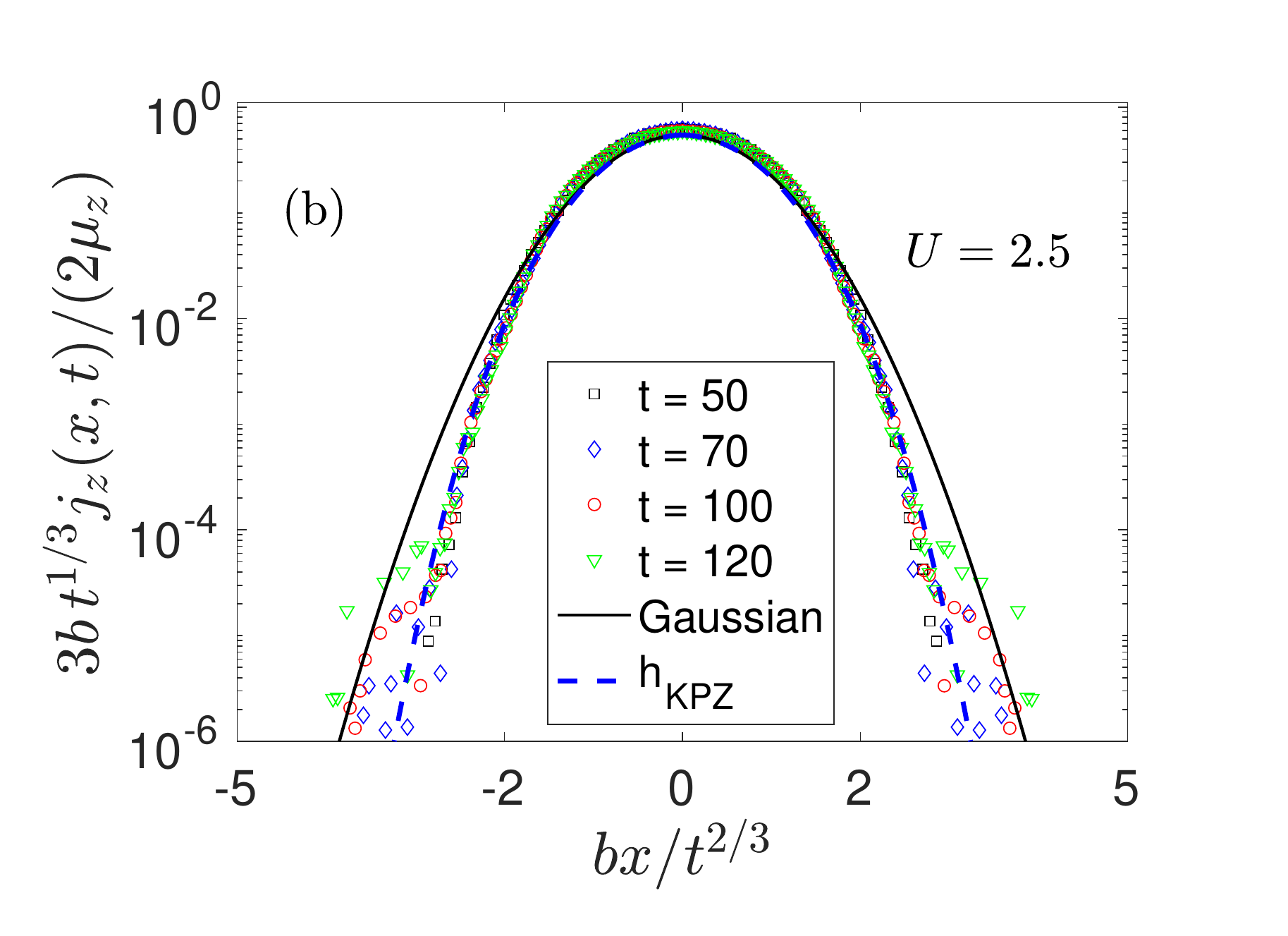}
  \caption{ 
  The 
  magnetization gradient, $\delta_x s_z(x,t) = \delta s_z(x,t) - \delta s_z(x-1,t)$
 (panel a), 
and the spin  current density $j^z(x,t)$ (panel b) display also universal KPZ scaling
at half filling. We use $b=0.98$ on both panels.
%\gzb{\bf Please, change axis label as for charge collapse. 
%We used  $\mu_z$ for the spin imbalance}
 }
  \label{fig:scaling_superdiffusive_m}
 \end{center}
\end{figure}

 We can thus determine the anomalous diffusion constant (in units of $a^2 \,J^{2\alpha}$)
by extracting the scaling factor $b$ from the current or density gradient profiles, 
and using the relation,  $D={\cal I}/b^2$. Alternatively, we can compute $\langle x^2\rangle$ by using the 
profile, $\delta_x n(x,t)$ as a probability distribution, and divide it by $t^{4/3}$. 
Both methods yield the same values  for $D$ within numerical precision. 

 The interaction dependence of $D$  is  displayed in Fig.~\ref{fig:diffusion_constant}.
Although we cannot assess the precise $U$ dependence of $D_c$ at small interactions,
 at moderate interactions  we find a scaling $D_c \sim U^{-1}$, 
which may be related to the reduction of the  hopping amplitude of doublons to neighboring empty sites, $\sim J^2/U$.
%\gz{\bf Pascu says he does not trust the large U data. We should  remove them. }

%but at larger interactions we observe clear deviations from this behavior.
% \textcolor{teal}{MAW: [\textbf{can we say anything about this? Why we expect $U^{-1}$? 
%Can we explain the deviation as some nonuniversal issue of our simulation (too short times?) }]}

\subsection{Spin sector}
%Using an analogous approach, 
We also analyzed the scaling  of the magnetization profile and that of the spin current density close to 
 the interface, and extracted the corresponding scaling functions.
All  our findings support the notion of   charge-spin duality at the $SU_c(2)\otimes SU_s(2)$ point. 
For example, in Fig.~\ref{fig:scaling_superdiffusive_m}, we show the scaling functions corresponding to
 the   magnetization gradient, $\delta_x s_z(x,t) $, and the spin current
$j_z(x,t)$. These obey  the  scaling equations, Eqs.~\eqref{eq:grad_n_scaling} and \eqref{eq:j_scaling},
but with $\delta_x n(x,t)$ and $ j(x,t)$ replaced by $\delta_x s_z(x,t)$ and $j_z(x,t)$, respectively.

\begin{figure}[b!]
 \begin{center}
  \includegraphics[width=0.85\columnwidth]{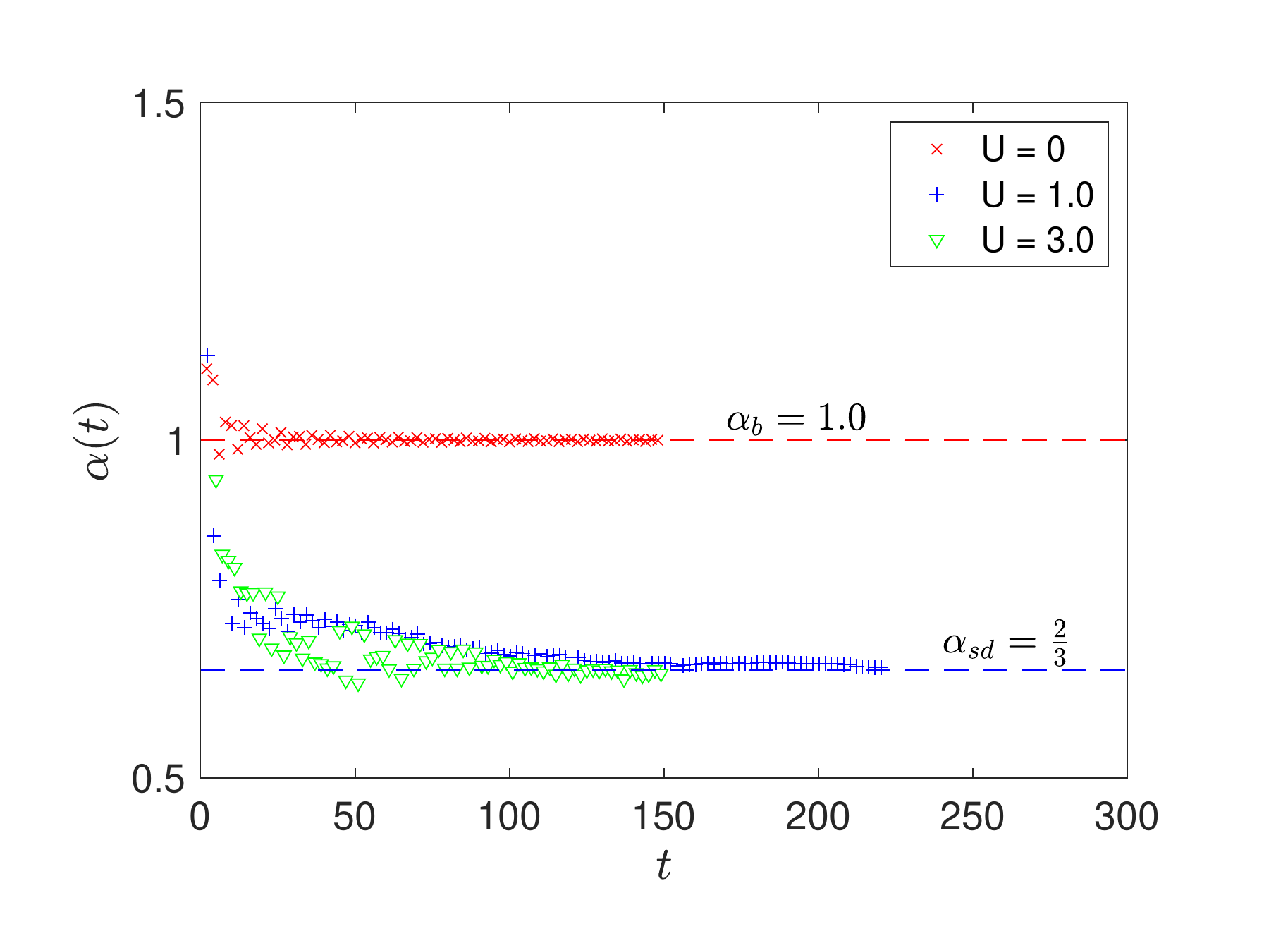}
  \caption{Scaling exponents obtained from the logarithmic differential  of the  charge transferred 
  across the interface. At $U=0$, the exponents approach the ballistic value, $\alpha_b = 1.0$ already at relatively short times. 
  At $U\neq 0$, the exponent converges to the asymptotic superdiffusive value, $\alpha_\text{sd} = {2\over 3}$. 
  For intermediate interactions (e.g., $U=1$), the convergence is slow and the asymptotic value is approached
  only  on time scales $t\approx 200$, which requires large system sizes (up to $L\approx 400$ sites). 
  For stronger interactions ( $U\ge 3$) the convergence is faster, and happens on time scales $t\lesssim100$.}
  \label{fig:scaling_exponents}
 \end{center}
\end{figure}

\subsection{Scaling exponents and finite size effects}
Our findings indicate quite clearly the presence of KPZ scaling.  However, obtaining  the scaling exponent accurately 
requires sufficiently large system sizes and a long enough  evolution times. To avoid spurious effects 
like reflection at the boundaries, the simulations' time span must be  restricted  by  the actual size of the system, 
$t \lesssim 2L$.
Additionally, the strength of the interaction is a crucial factor to consider.
Fig.~\ref{fig:scaling_exponents} displays the exponent $\alpha = 1/z$ obtained from the logarithmic differential
of the total charge across the interface, 
\be 
N_\text{tr}(t) \sim t^{\gzb{\alpha}}\;.
\ee
In the ballistic case, $U=0$, the convergence to the asymptotic values occurs rapidly, and relatively small system sizes are sufficient for accurate calculations. On the other hand, for $U\ne 0$, convergence depends crucially on the interaction strength; 
 for stronger  interactions the logarithmic differential converges faster, although it becomes more noisy, 
 while for small interactions convergence is very slow. 
\begin{figure}[t!]
 \begin{center}
  \includegraphics[width=0.85\columnwidth]{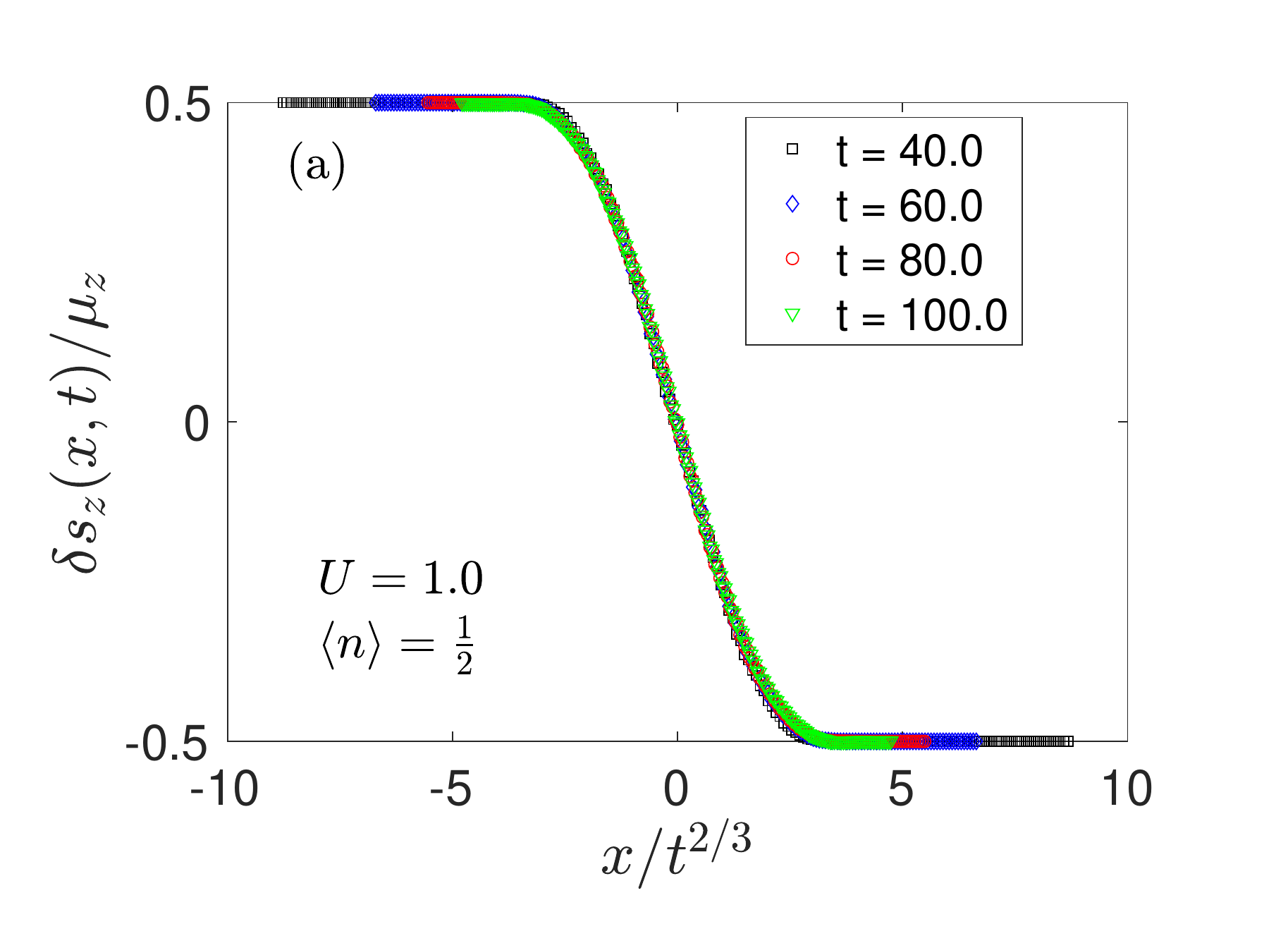}
  \includegraphics[width=0.85\columnwidth]{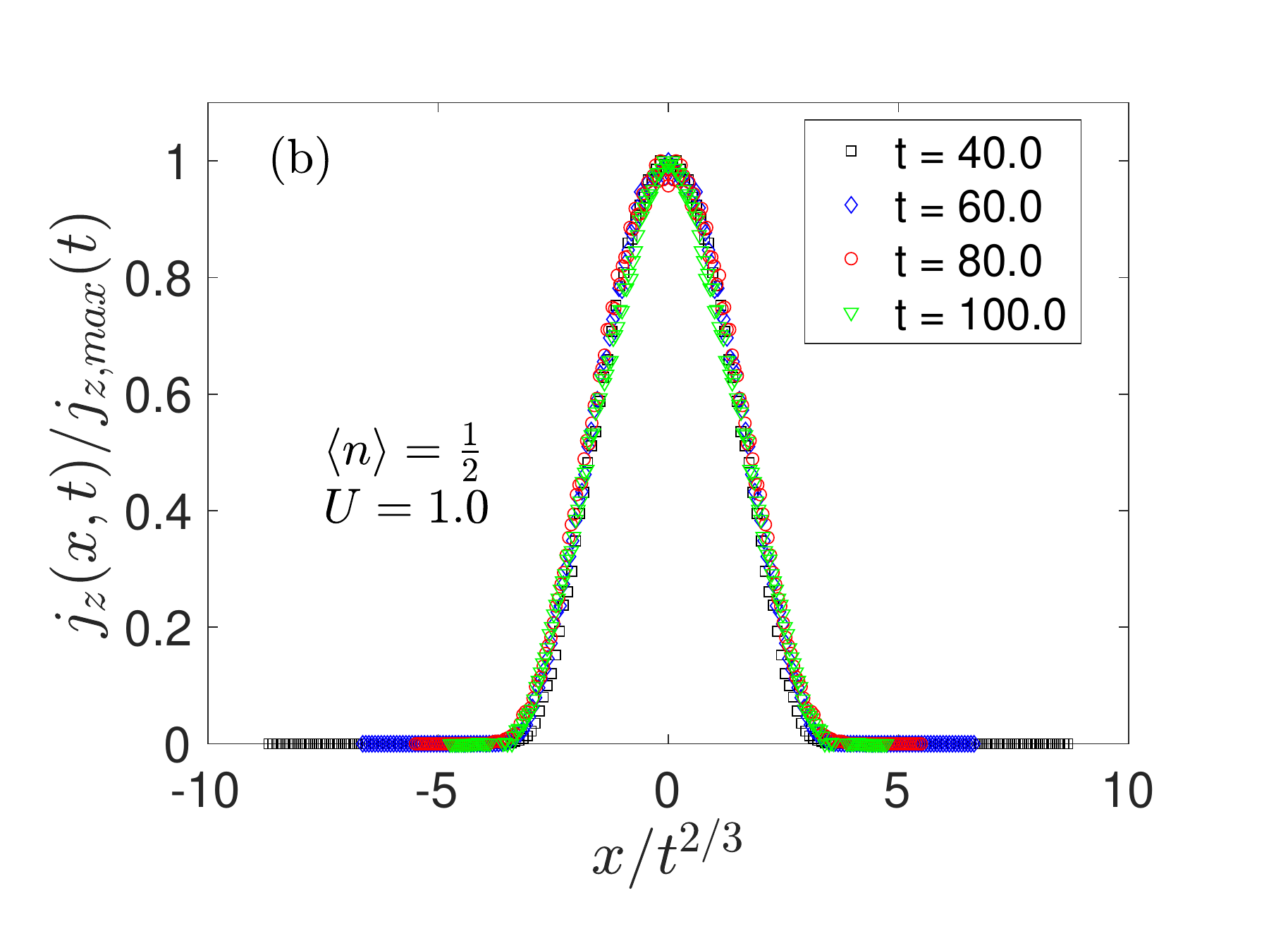}
  \caption{The rescaled profiles of (a) the average magnetization $\delta s_z(x,t)$ at a quarter filling, where $\langle n\rangle=1/2$, and (b) the average spin current $j_z(x,t) $ along the chain, both exhibiting $x/t^{2/3}$ scaling.}
  \label{fig:scaling_away_hf}
 \end{center}
\end{figure}
\begin{figure}[b!]
 \begin{center}
  \includegraphics[width=0.8\columnwidth]{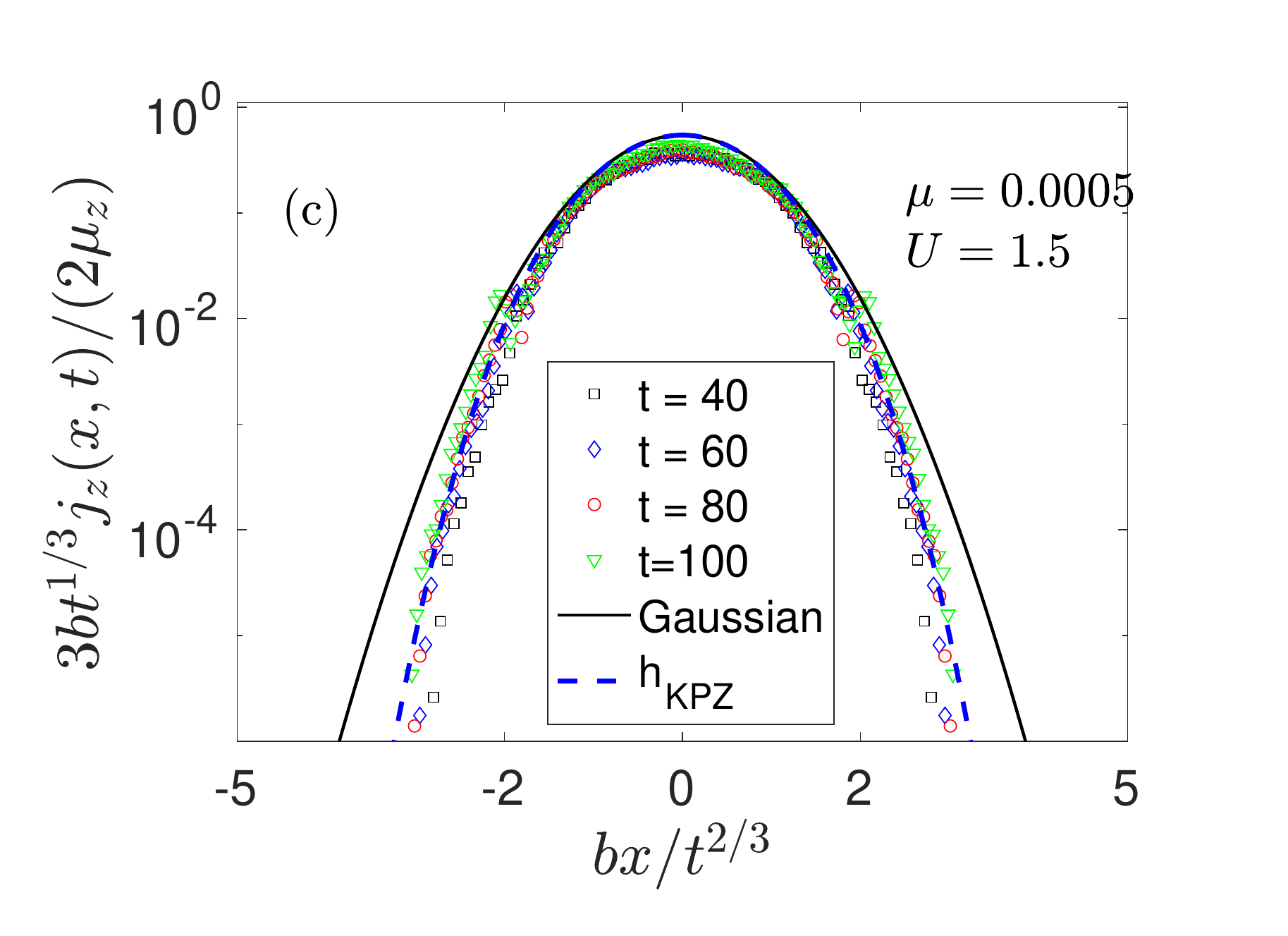}
  \caption{The scaling function for the spin current along with the corresponding KPZ scaling function at quarter filling $\average{n}=1/2$ and $U=1.5$. The parameter $b$ to a fixed to $b=0.75$.}
  \label{fig:universal_fct_away_hf}
 \end{center}
\end{figure}
 Even for  intermediate values, $U\sim 1$,  time scales of different processes overlap, 
 and obtaining the correct exponent 
 requires simulations of very long  duration and, correspondingly,  large enough system sizes.
 Although we cannot reach convergence for $\text{d}\ln(N_\text{tr}(t))/ \text{d}\ln\,t $
  in  the small coupling regime $U\lesssim 1$, the tendency 
 observed at intermediate couplings brings us to conclude that at sufficiently long times 
 a \emph{superdiffusive} behavior   emerges at any finite coupling, $U$. 
 As we shall see, 
sufficiently large system sizes and  a very careful analysis is needed also the open system setup, to investigate the asymptotic behavior.

\begin{figure}[t!]
 \begin{center}
  \includegraphics[width=0.9\columnwidth]{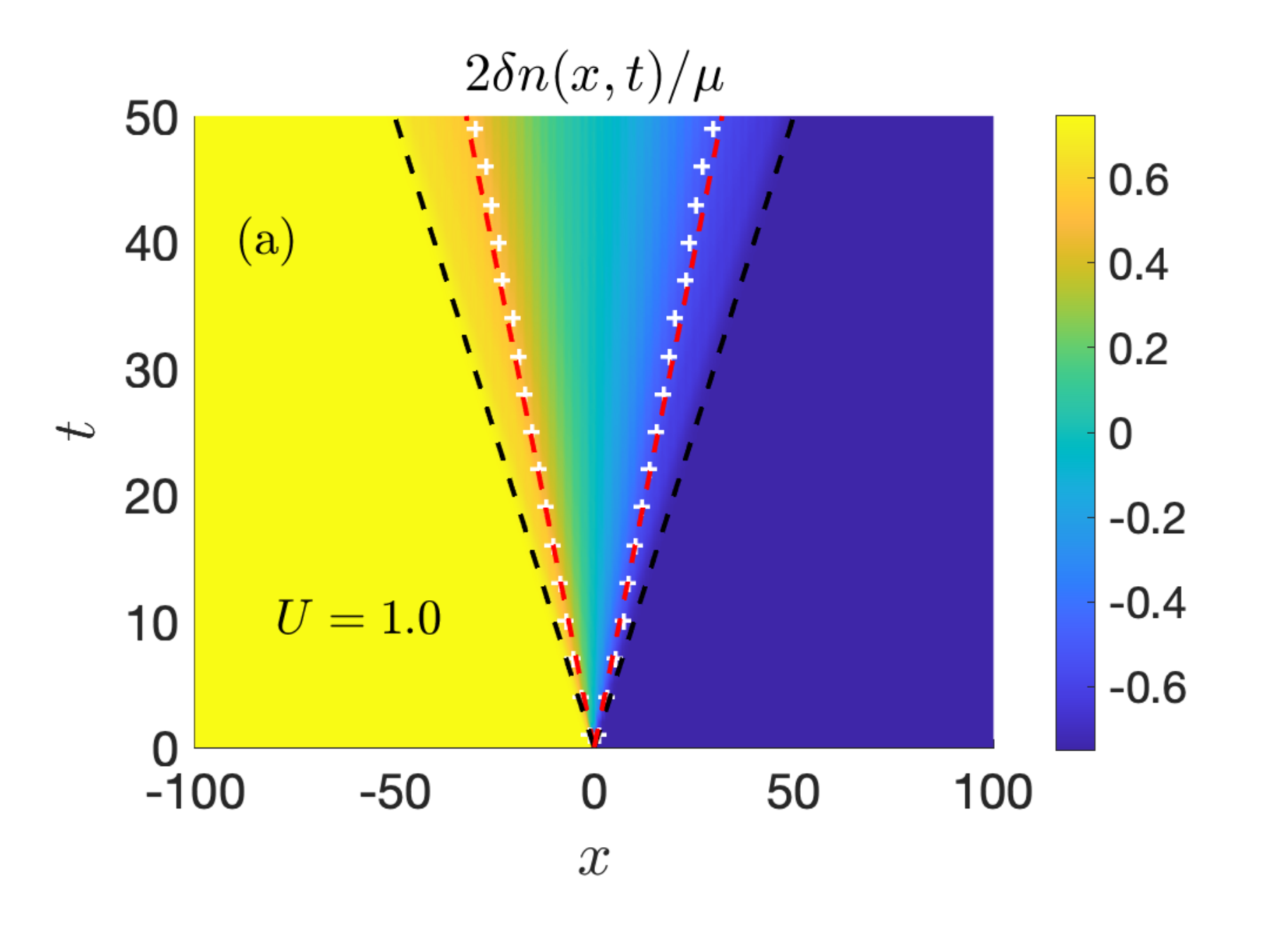}
  \includegraphics[width=0.9\columnwidth]{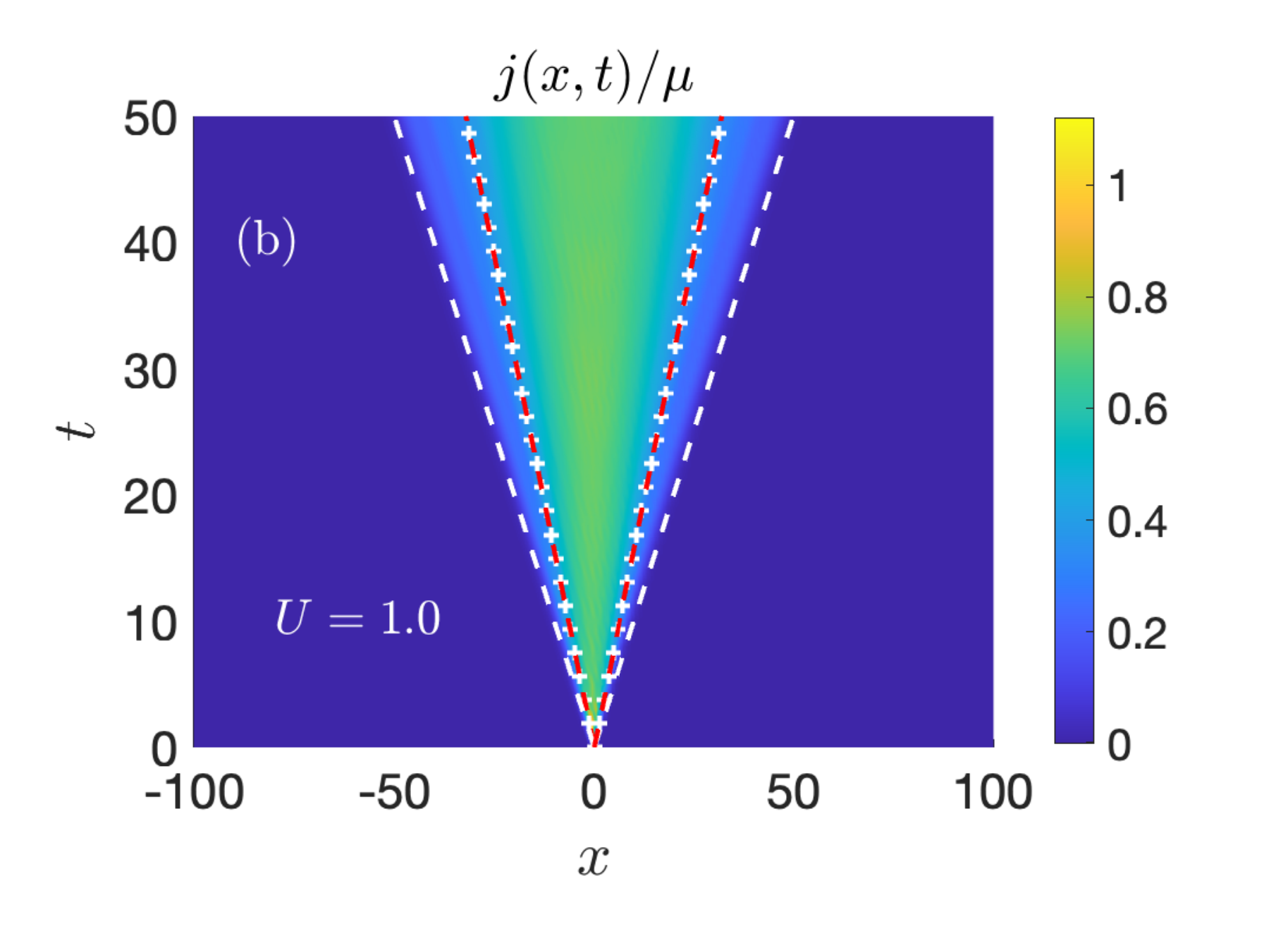}
  \caption{Time evolution of the average occupation and average charge current along the chain at quarter filling for $U=1.0$. Two fronts moving at different velocities are crearly visible.}
  \label{fig:away_hf_n}
 \end{center}
\end{figure}

%%%%%%%%%%%%%%%%%%%%%%%%%%%%%%%%%%%%%%%%%%%%%%%%
\section{Away from half-filling ($U\ne 0$, $\average{n}\ne 1$)}\label{sec:away_half_filling}
%%%%%%%%%%%%%%%%%%%%%%%%%%%%%%%%%%%%%%%%%%%%%%%%

Away from half-filling, the $SU_c(2)$ pseudo-charge symmetry is reduced to a $U_c(1)$ symmetry.
This implies that the density matrix $\rho(t)$ is no longer invariant under  rotations generated by $\boldsymbol \eta$. 
On the other hand, the model still retains its spin $SU_s(2)$ symmetry. 
%since the total spin, as represented by the operator $\mathbf S$, remains a good quantum number.
%With the $SU_s(2)$ symmetry intact, 
We are thus able to investigate the conjecture that links non-abelian symmetries and the KPZ scaling
in the spin sector, even when the $SU_c(2)$ symmetry is broken.

In the following, we present data  both in  the spin and in the charge sectors, obtained at quarter filling, %where the average electron density is
 $\average{n}=1/2$.
Fig.~\ref{fig:scaling_away_hf} shows the scaling behavior of the magnetization and spin current profiles, along with the corresponding scaling function.
%Our findings reveal that 
Only a scaling with $x/t^{2/3}$ results in a proper data collapse for both $\delta s_z(x,t)$ and $ j_z(x,t)$, and the universal curve displayed in Fig.~\ref{fig:universal_fct_away_hf} demonstrates a distinct KPZ dependence.
      
\begin{figure}[t!]
 \begin{center}
  \includegraphics[width=0.9\columnwidth]{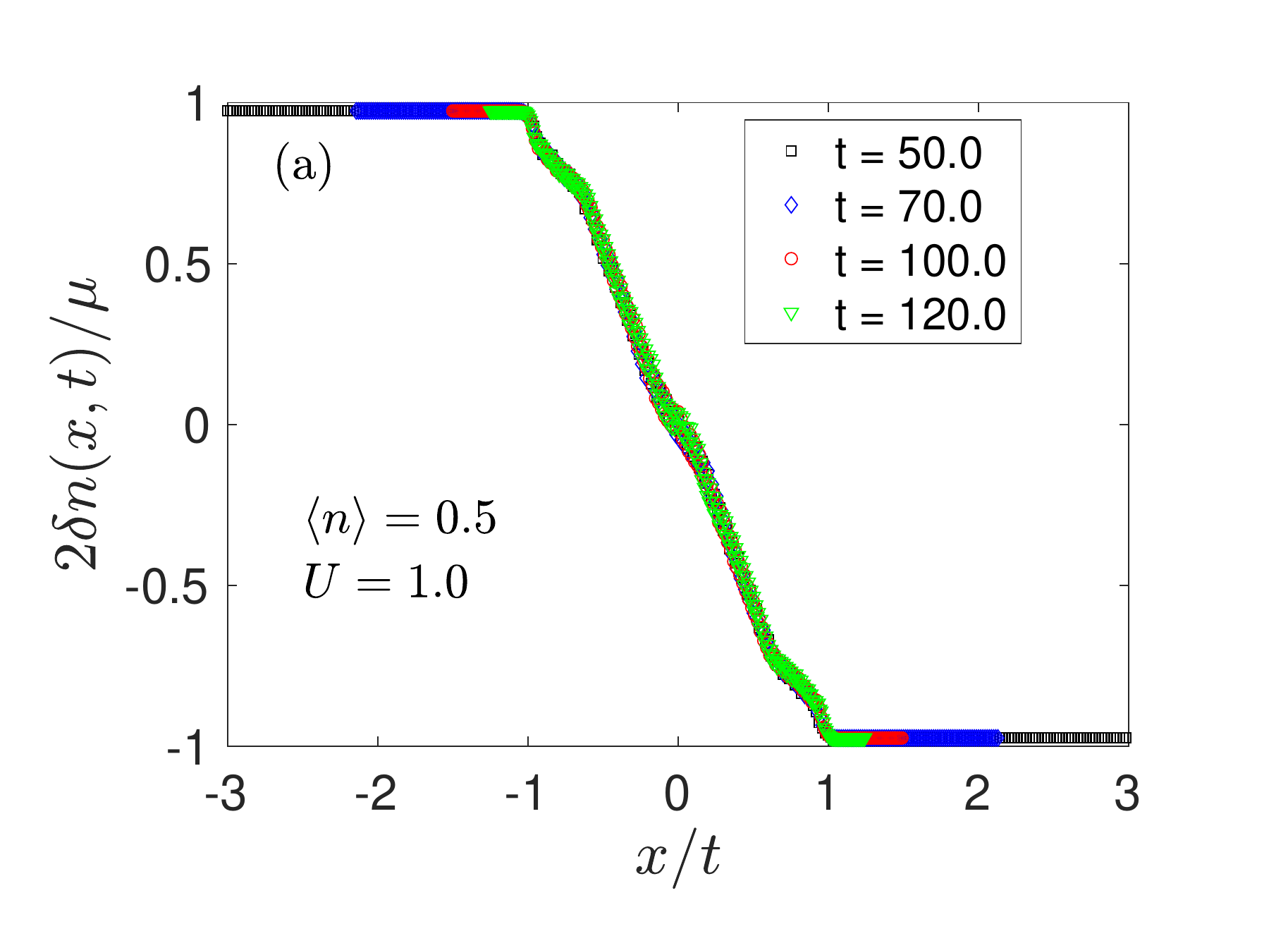}
  \includegraphics[width=0.9\columnwidth]{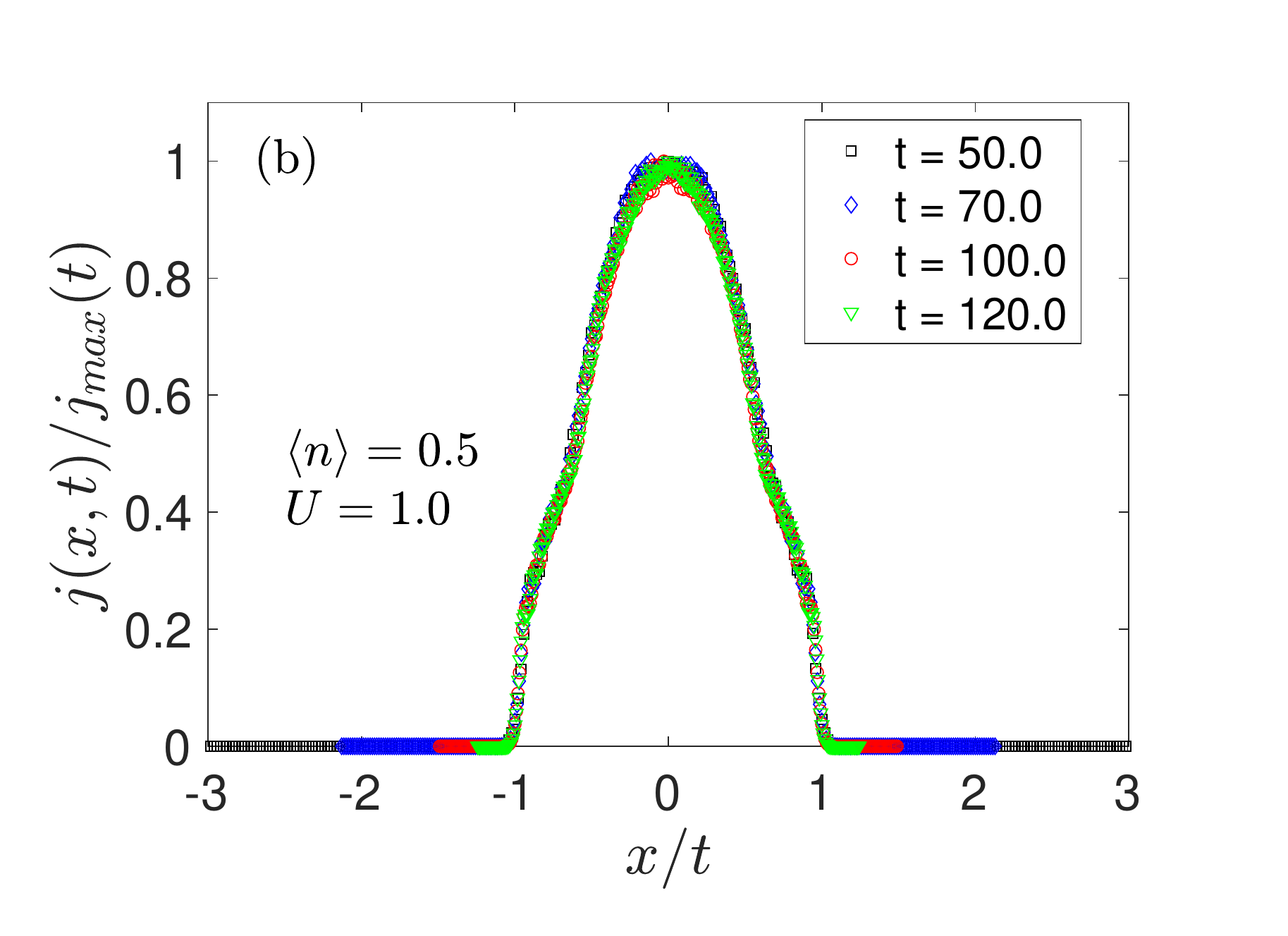}
  \caption{The rescaled profiles for the average occupation (a) and current (b) at quarter filling for $U=1$ display ballistic scaling.}
  \label{fig:away_hf_n_cuts}
 \end{center}
\end{figure}

In contrast, quasi-particles move ballistically in the charge sector. 
The average occupation $\delta n(x,t)$, e.g., presented in Fig.~\ref{fig:away_hf_n} 
displays a typical light cone pattern. 
Interestingly, the occupation profile  shows two distinct velocities: one that corresponds to the maximum possible velocity of the free particles, i.e., 
$v_\text{max}=J$, and another, smaller velocity that depends on the strength of the interaction $U$. These findings are consistent with similar observations in the context of GHD, where multiple light velocities are also observed in some cases~\cite{Piroli.2017,Nozawa.2020,Nozawa.2021}. Figs.~\ref{fig:away_hf_n_cuts}.a and b
show the spatial charge and charge current profiles at various times and demonstrate that only a ballistic rescaling, $\sim x/t$, 
result in the proper data collapse for both the  occupation and the  charge current, and also  the clear development of the two fronts.

%%%%%%%%%%%%%%%%%%%%%%%%%%%%%%%%%%
\section{Hubbard chain coupled to two reservoirs}\label{sec:wl}
%%%%%%%%%%%%%%%%%%%%%%%%%%%%%%%%%%

\begin{figure}[b!]
 \begin{center}
  \includegraphics[width=0.9\columnwidth]{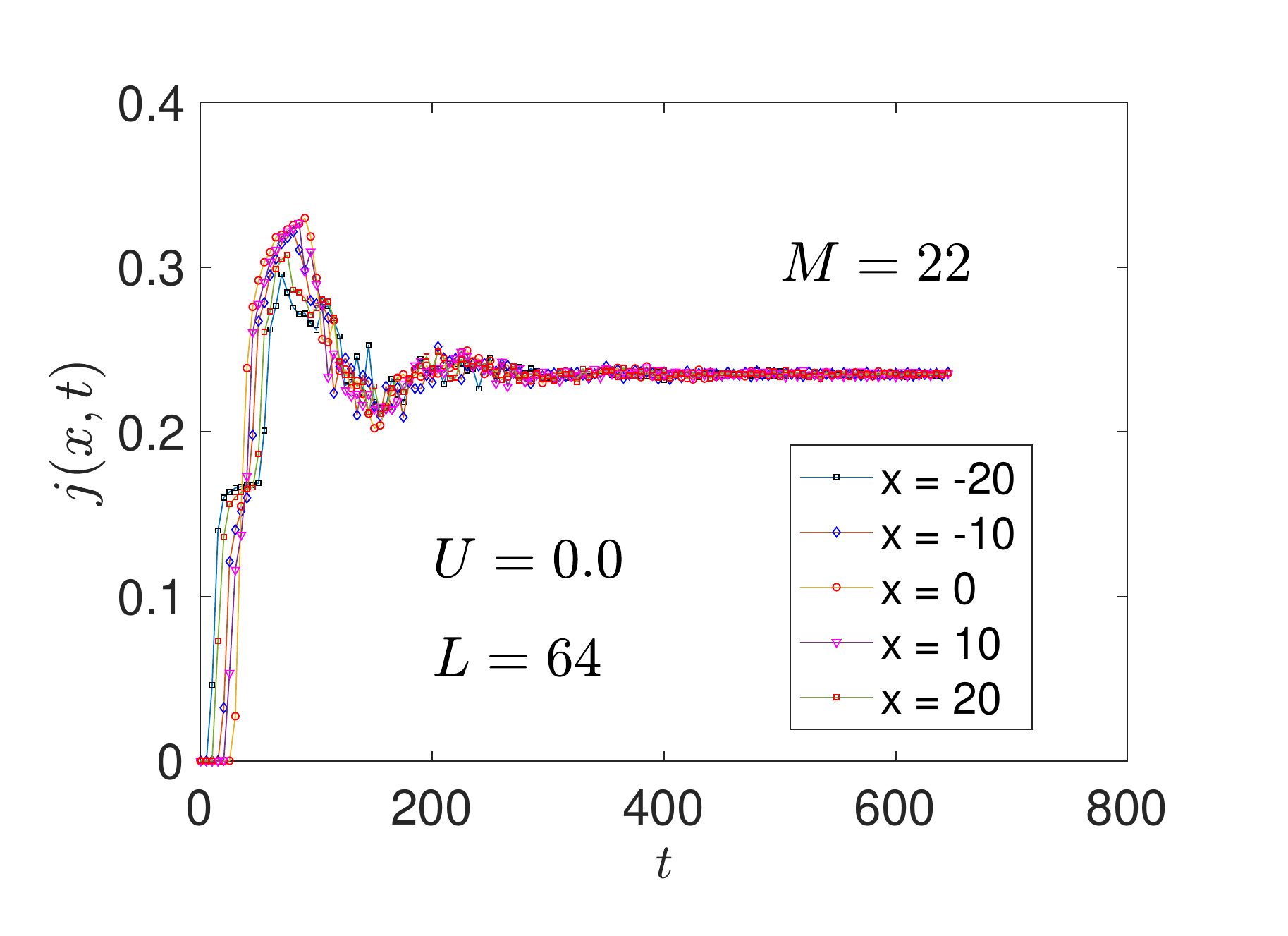}
  \includegraphics[width=0.9\columnwidth]{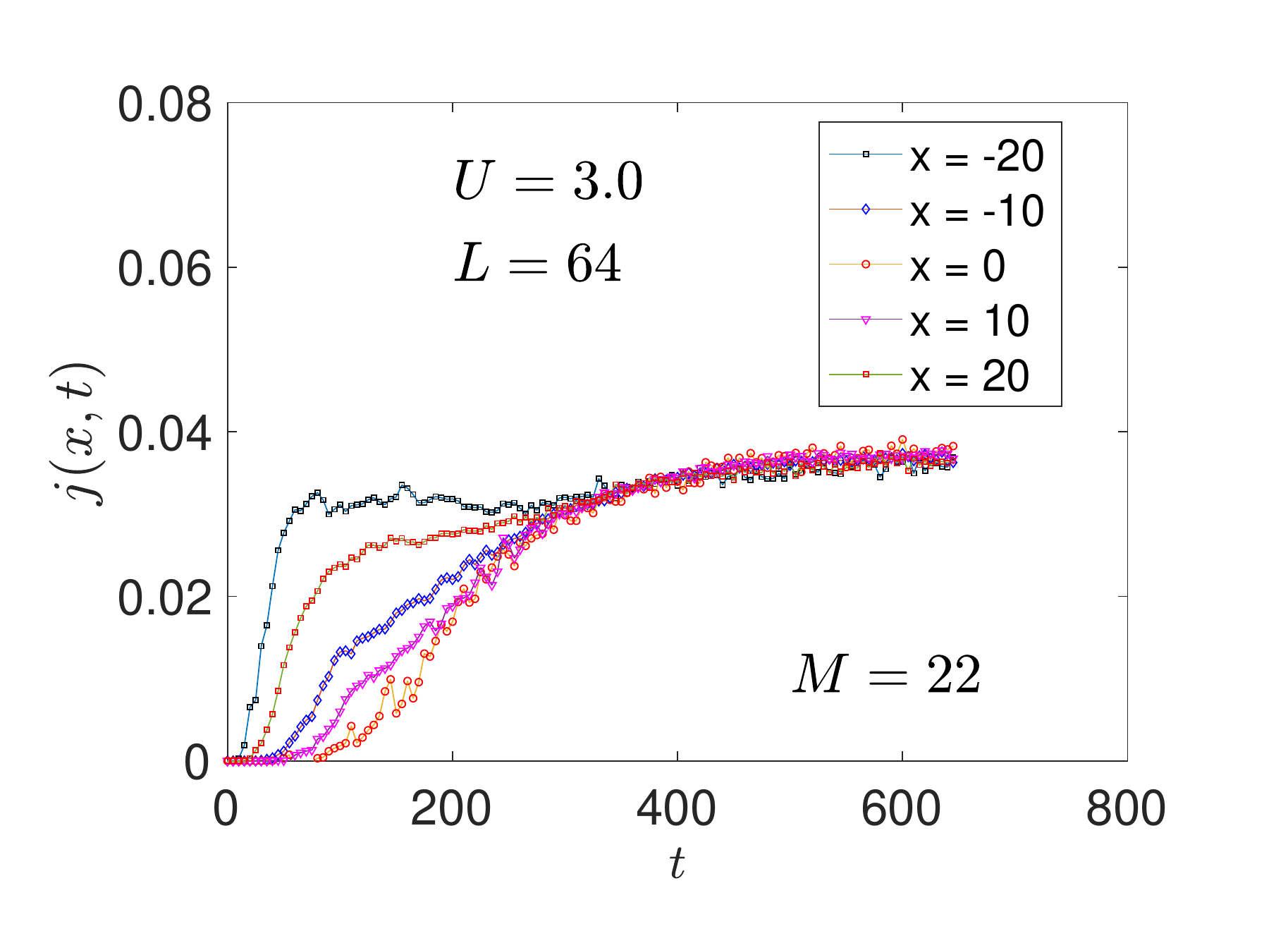}
  \caption{The evolution of the current as a function of time towards the NESS is shown for different bond positions $x$
  along the chain
  for (a) the non-interacting case ($U=0$) and (b) $U=3.0$.
  The chain length is fixed to $L=64$ sites. In the interacting regime, the system evolves  more slowly towards the NESS.}
  \label{fig:wl_current}
 \end{center}
\end{figure}

\begin{figure}[t!]
  \begin{center}
  \includegraphics[width=0.8\columnwidth]{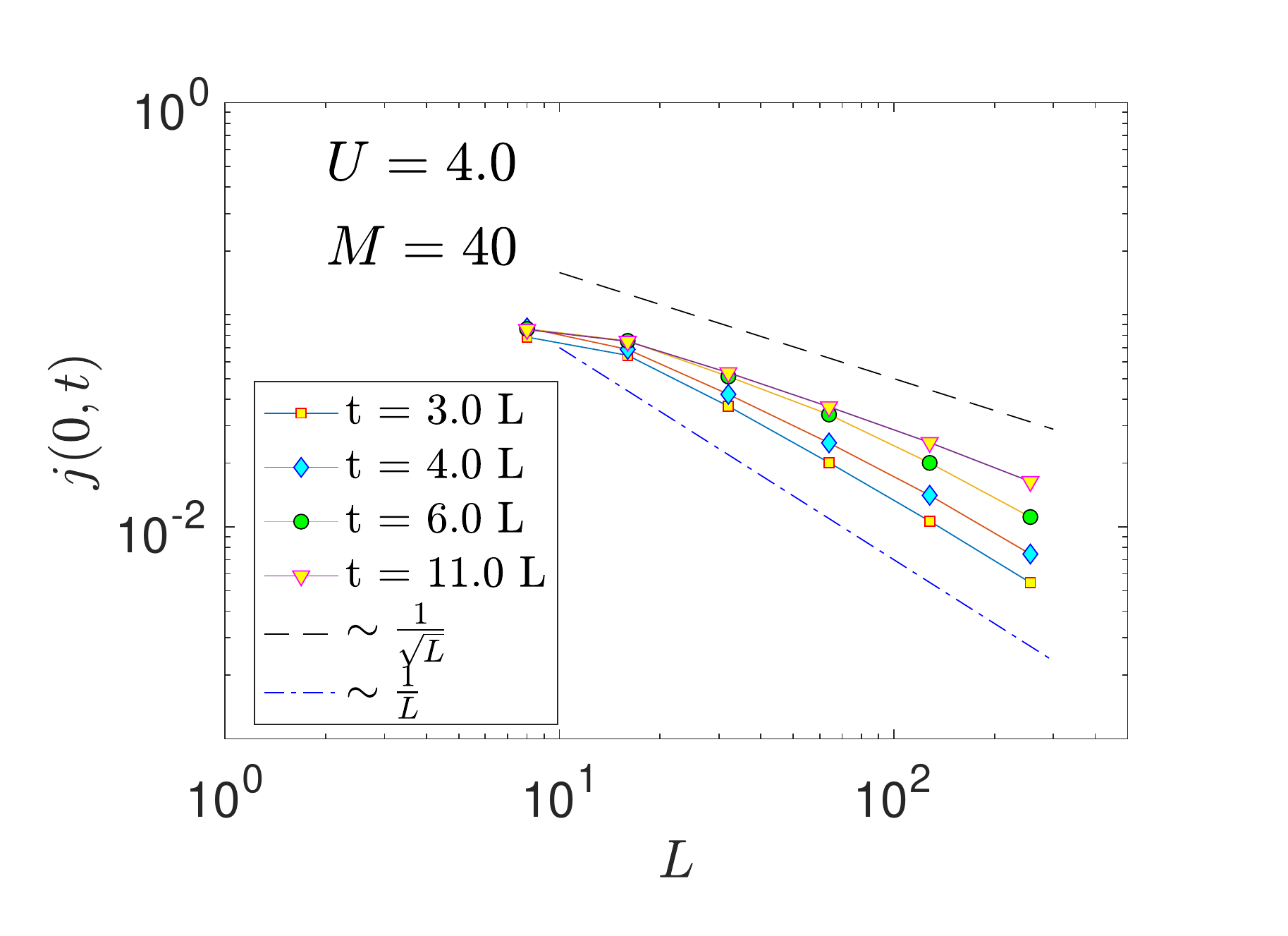}
  \caption{Scaling of the current  as function of the chain length for $U=4.0$ at different times.} 
\label{fig:wl_time}
%  \label{fig:current_scaling_time_dependence}
 \end{center}
\end{figure}

So far we have investigated KPZ scaling in terms of a quench protocol. 
We now turn to a different setup, and 
 explore the emergence of the KPZ scaling in a Hubbard chain  coupled to external {markovian} (Lindbladian) reservoirs, 
 gerenating local particle gain and loss at both ends.
We investigate the dynamics and various observables, such as the average occupation along the chain and the average 
current in the non-equilibrium steady state (NESS).

We perform TEBD calculations~\cite{Vidal.2003, Vidal.2004} by starting from an infinite temperature state, 
and solving the Lindlbad equation
\begin{equation}
 i\dot \rho =\cL [\rho] = [H, \rho] + i \cD[\rho]\label{eq:Lindblad}
\end{equation} 
in the Gorini, Kossakowski, Sudarshan~\cite{Gorini.1976} and Lindblad  approach~\cite{Lindblad.1976}.
Here $H$ is the Hamiltonian introduced in Eq.~\eqref{eq:Hubbard}, while the dissipator term $\cD[\rho] = \sum_F \Gamma_F \cD_F[\rho]$
\begin{equation}
 \cD_F[\rho] = 2F\rho F^\dagger -\{F^\dagger F, \rho\} 
\end{equation}
is given in term of the boundary jump operators 
$F \in \{c_{-{L/2}\sigma}, c^\dagger_{-{L/ 2}\sigma}, c_{{L/ 2}\sigma}, c^\dagger_{{L/}\sigma} \}$, acting at the first and last sites of the chain.

\begin{figure}[b!]
 \begin{center}
  \includegraphics[width=0.85\columnwidth]{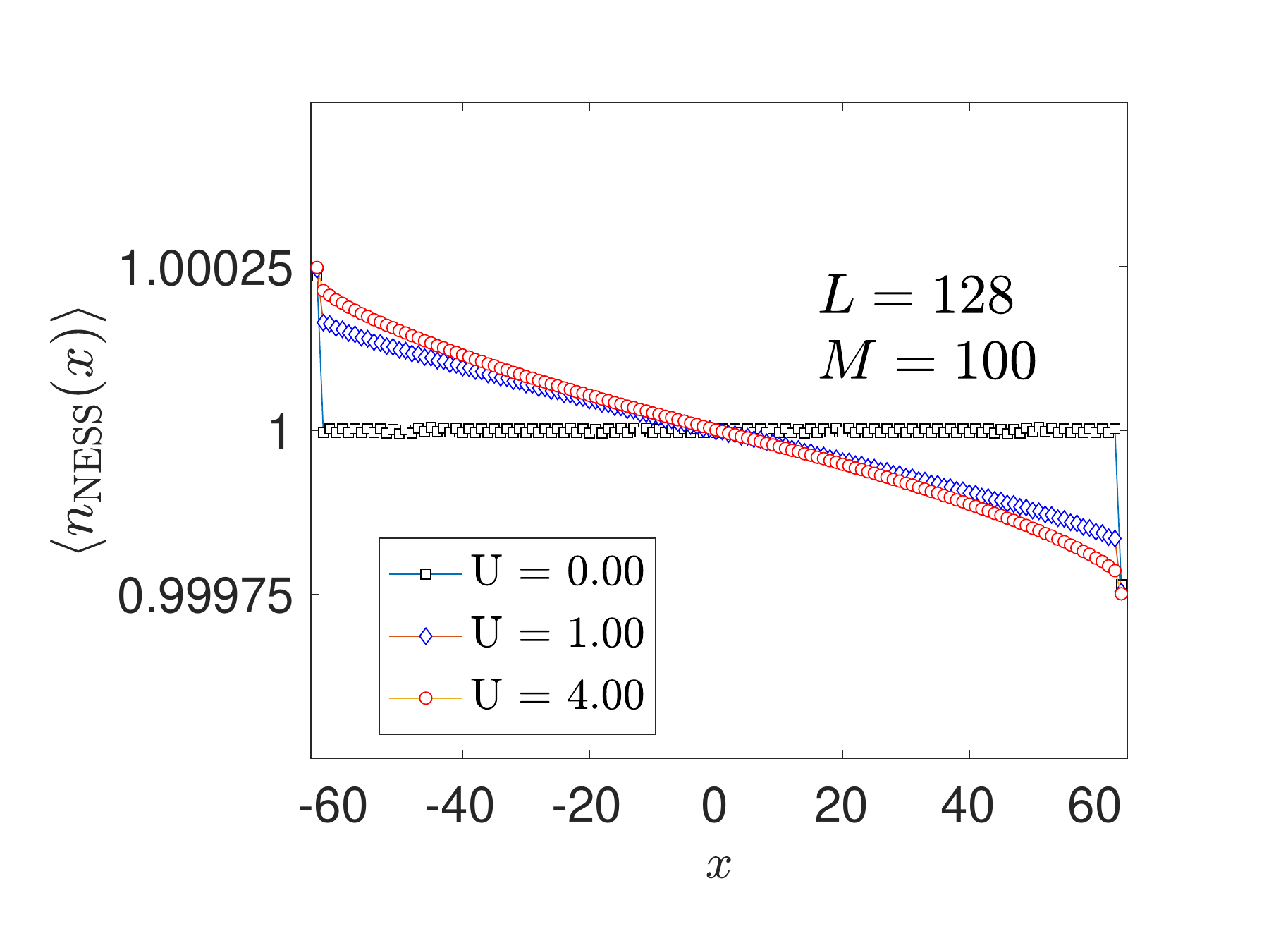}
  \caption{ Average occupation along the chain in the NESS for a chain of length $L=128$ and different values of $U$.
}
  \label{fig:wl_occupation}
 \end{center}
\end{figure}

%%%%%%%

To induce a non-equilibrium steady state (NESS) along the chain, we introduce a slight imbalance in the couplings 
for  loss and gain processes  at the two ends of the chain. Specifically, on the left % at one end, the coupling
the strength for the loss process, 
$\Gamma^{(-)}_{-{L/ 2},\sigma}= \Gamma \,(1-\mu)$ is less than that of the  gain process,
$\Gamma^{(+)}_{-{L/ 2},\sigma}={\Gamma}\,(1+\mu)$ for $\mu >0$, 
while on the right, loss processes are increased and gain processes reduced,
 $\Gamma^{(\pm)}_{L/ 2,\sigma} =  \Gamma^{(\mp)}_{-L/ 2,\sigma}$. 
 %Here we follow the convention that \gz{$\Gamma^\pm$}, denote  couplings to  the annihilation
 %and  creation operators, respectively. 
   Here $\Gamma$
controls the strength of the coupling to the reservoirs, while  the imbalance $\mu\ll 1$ plays the role of an external chemical potential that drives the system 
toward a NESS.

We used the NA-TEBD approach  of Ref.~\cite{Moca.2022} to integrate the Lindblad equation~\eqref{eq:Lindblad}.
We also verified that  the resulting NESS is unique and independent of the initial conditions. 
Using the infinite temperature state 
as the initial state, however, allows us to reach  NESS more quickly. 
We presume  that NESS is reached once the average 
current $ j(x,t)$ no longer depends on the position $x$ and time. Fig.~\ref{fig:wl_current} 
shows the evolution of the current at several bond positions along the chain, 
indicating that stronger interactions lead to slower dynamics and a slower convergence towards 
 NESS.

To  capture the KPZ scaling  accurately, it is necessary to evolve the system for sufficiently 
long times,  $t\approx 10L/v_F$, such that the system reaches its non-equilibrium steady state. 
Figure~\ref{fig:wl_time}  presents the combined finite size and time  dependence   of the  current 
at the center, $j(x=0,t)$. At short times, the current scales in a diffusive manner,  
$j(0,t)\sim 1/L$~\cite{Prosen.2012}, but displays a superdiffusive scaling,   $j(0,t) \sim 1/\sqrt{L}$, 
once  simulation times are long enough.

\begin{figure}[t!]
  \begin{center}
   \includegraphics[width=0.9\columnwidth]{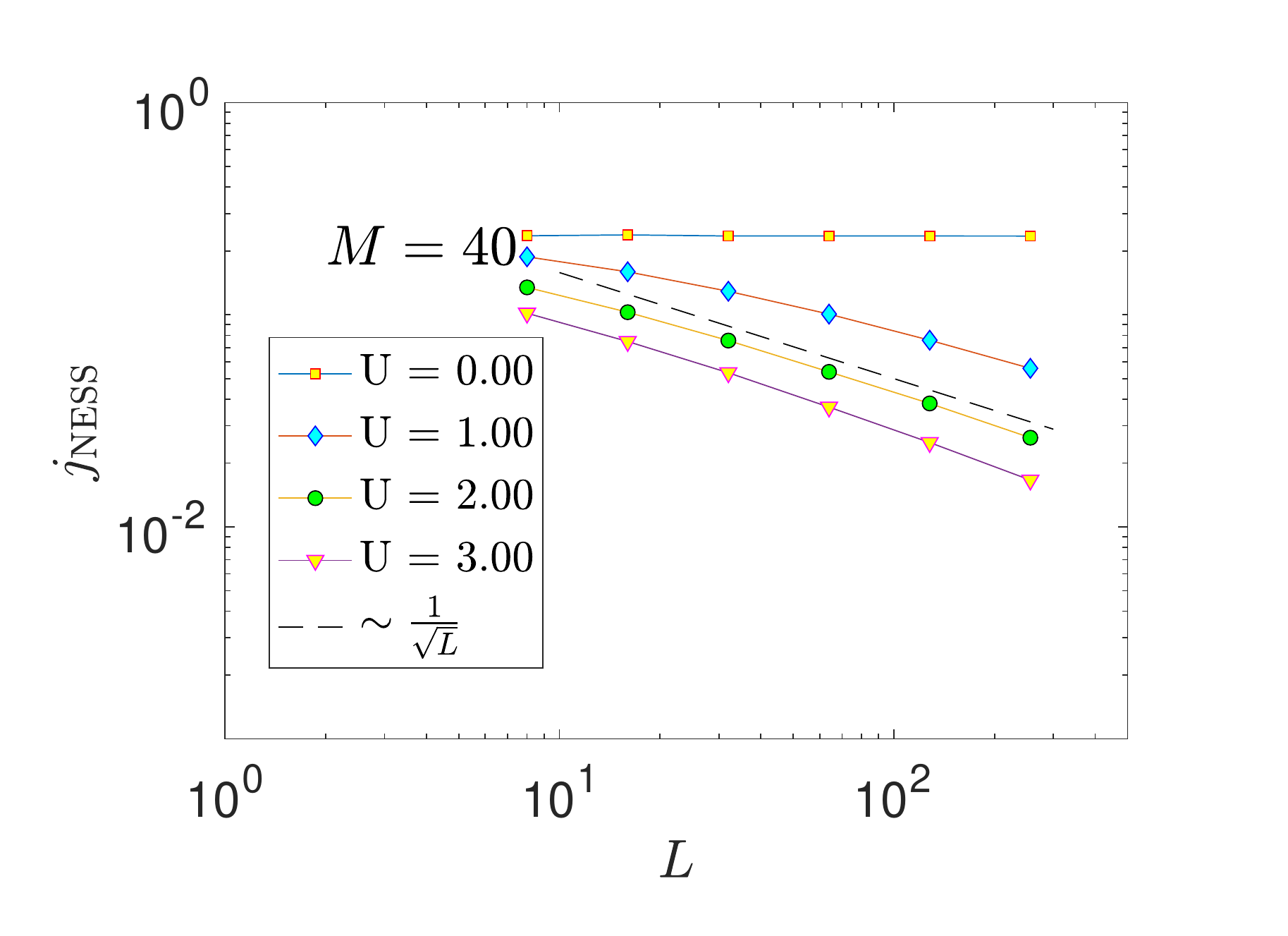}
  \includegraphics[width=0.85\columnwidth]{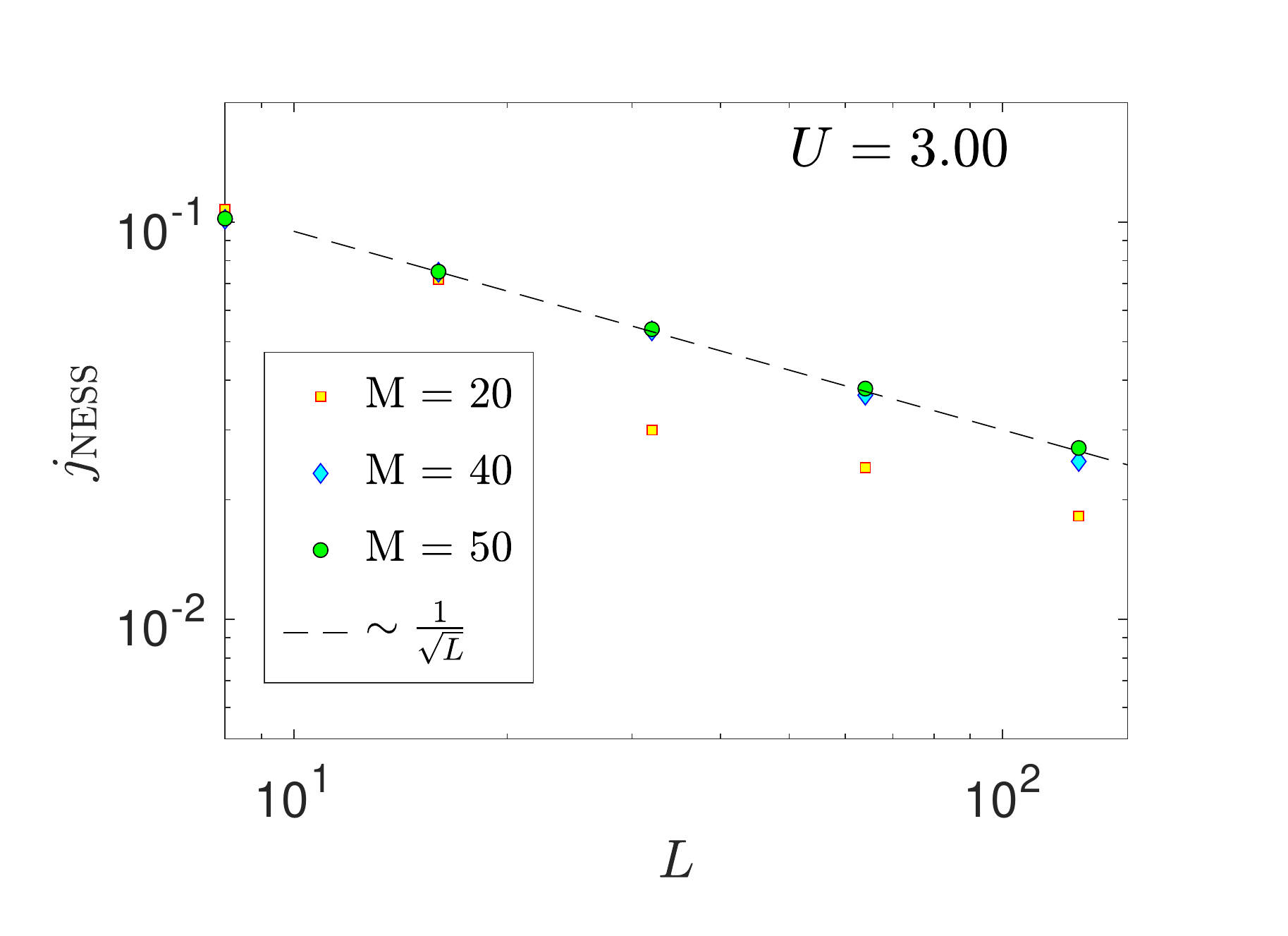} 
  \caption{Top:
  Interaction and size  dependence of $j_{\rm NESS}$.
  For $U=0$, the current is constant, indicating a ballistic behavior, while for finite $U$
  $j_{\rm NESS}\sim {1\over \sqrt{L}}$, confirming  KPZ scaling.
   Bottom:   Bond dimension dependence  of the steady state current in the middle of the chain as a function of the chain length
   for $U=3.0$.} 
   %
%  \label{fig:wl_time}
  \label{fig:current_scaling}
 \end{center}
\end{figure}

Fig.~\ref{fig:wl_occupation} depicts a typical occupation profile along the chain in the NESS. At $U=0$, 
the average occupation is approximately $1$, except for the first and last sites, which exhibit a jump. 
%of around $\pm\mu/2$. 
We have validated these $U=0$ TEBD results  against the $3^{rd}$ quantization
approach~\cite{Prosen.2008} (not shown). As $U$ increases, the average occupation is no longer uniform along the chain, 
but there is still a small jump at the boundaries. 
%typically much smaller than $\mu/2$.

%
% \begin{figure}[t!]
%  \begin{center}
%  \includegraphics[width=0.9\columnwidth]{WL_current_scaling_mu_0.0005_Mmax_40.pdf}
%  \caption{ Scaling of the current in the middle of the chain as function of the chain length.
%  For $U=0$, the current is constant, indicating a ballistic behavior, while for finite $U$ the
%  the scaling $j_{\rm NESS}\sim {1\over \sqrt{L}}$ confirms the KPZ scaling. }
%  \label{fig:wl_scaling_current}
% \end{center}
%\end{figure}
%

%
%
%\begin{figure}[t!]
% \begin{center}
% \includegraphics[width=0.9\columnwidth]{WL_current_scaling_mu_0.0005_U_3.00.pdf}
% \caption{ Scaling of the steady state current in the middle of the chain as a function of the chain length for $U=3.0$, for different bond dimensions. }
% \label{fig:wl_scaling_current_bond_M}
% \end{center}
%\end{figure}

Fig.~\ref{fig:current_scaling} illustrates how the  NESS current, $j_{\rm NESS}$, 
changes with  system size. For $U=0$, the current remains constant regardless of the system size, $j_{\rm NESS}\sim 1/L^0$,
characteristic of   ballistic transport.
In contrast, for  $U\ne0$, $j_\text{NESS}$
exhibits a power law dependence on the system size,
most consistent with $j_\text{NESS}\sim 1/\sqrt{L}$, indicative of superdiffusive transport~\cite{Znidaric.2011}. 
%his behavior corresponds to a KPZ scaling and is consistent with the geometry of the quench protocol discussed in previous sections. 

In Fig.~\ref{fig:current_scaling} we also display the bond dimension dependence of the NESS current. 
%for different bond-dimensions. 
Although for small system sizes the slow growth of operator entropy allows us to 
use small bond dimensions $M\approx 20$,  for  system sizes exceeding $L>100$ it is necessary to raise the bond 
dimension beyond $M>100$ to obtain reliable results and  to capture the scaling exponents accurately, 
especially for larger values of $U$.

\section{Conclusions}\label{sec:conclusions}

In this work, we have investigating  KPZ scaling in the Hubbard model, and explored its  connection with 
non-abelian symmetries. In this regard, the Hubbard model provides an ideal playground, since 
it is integrable, and at half-filling it possesses two non-abelian symmetries, which can both be broken. 
%We chose our Hamiltonian based on its rich non-abelian symmetries and integrability. }

For our numerical analysis, we utilized two setups. In the first, 'quench' setup we used a quench protocol, 
whereby we prepared  the system 
in a mixed state asymptotically close to the $T = \infty$ state, but  with  weakly imbalanced 
spatial occupation or magnetization profiles. In the
second, 'open' setup we attached particle sources and drains to drive current through the 
system, and investigated the finite size scaling of the  steady state (NESS) current in the linear response regime.

Our calculations confirm  that, at half-filling, any small interaction induces KPZ scaling for spin and charge transport:
 both charge and spin  profiles show a distinctive $x/t^{2/3}$ collapse. 
 Our accuracy allowed us to extract the density-density correlation's  function, 
 $\langle n(x,t)n(0)\rangle$ from the data,  and confirm that  its scaling properties are captured with 
 the universal KPZ  scaling function within our numerical accuracy. This scaling analysis also enabled us to extract the 
 anomalous diffusion constant, $D$ and its interaction dependence. The anomalous diffusion  
 is suppressed at large values of $U$, and $D$ seems to vanish in the $U\to \infty$ limit. This is somewhat counterintuitive, 
 since, as discussed below, charge transport becomes ballistic once charge $SU_c(2)$ symmetry is broken, 
 and charge fronts can propagate with the  Lieb-Robinson velocity, $v_\text{max} = a\,J$.
 
As mentioned above,  charge transport becomes  once we break the charge $SU_c(2)$
 symmetry down to $U_c(1)$, by moving away from half-filling.  
 In this case a double front structure appears, associated with two distinct front velocities; the first front 
 propagates with $v_\text{max} = a\,J$, while the second front legs behind  with a smaller, interaction dependent velocity. 
 Spin transport remains, however superdiffusive, as apparently implied by the unbroken $SU_s(2)$
 symmetry. The scaling of the NESS current in the open geometry confirmed the superdiffusive scaling as well.

From these calculations it is thus clear that superdiffusive behavior is associated with non-abelian symmetries. 
Whether integrability is a crucial ingredient, needs to be investigated, though semiclassical results seem 
to imply that integrability is not essential for the KPZ scaling observed~\cite{Moessner_PRB_2022}. 

\section*{Acknowledgments}
We thank \" Ors Legeza for insightful discussions.
This research is supported by the National Research, Development and Innovation 
Office - NKFIH through research grants Nos. K134983 and SNN139581, within
the Quantum National Laboratory of Hungary. 
C.P.M acknowledges support by the Ministry of Research, Innovation and Digitization, CNCS/CCCDI–UEFISCDI, 
under projects number PN-III-P4-ID-PCE-2020-0277, under the project for funding the excellence, contract No. 29 PFE/30.12.2021.
T.P. acknowledges ERC Advanced grant 694544-OMNES and ARRS research program P1-0402.
M.A.W has also been supported by the Janos Bolyai Research Scholarship of the
Hungarian Academy of Sciences and by the ÚNKP-22-5-BME-330 New National Excellence Program of the Ministry for Culture and Innovation from the source of the National Research, Development and Innovation Fund. 
We acknowledge KIF\"U for awarding us access to resource based in Hungary.

\bibliography{references}

\end{document}